\documentclass[prl,twocolumn,aps,superscriptaddress,longbibliography]{revtex4-2}
\usepackage[T1]{fontenc}
\usepackage[utf8]{inputenc}
\usepackage{refstyle}
\usepackage{amsmath}
\usepackage{amssymb}
\usepackage{graphicx}
\usepackage[pdftex, hidelinks]{hyperref}
\usepackage[usenames,dvipsnames]{xcolor}
\usepackage{bbm}
\usepackage{tikz}
\usepackage{bm}
\usepackage{babel}
\usepackage[makeroom]{cancel}
\usepackage{comment}
\usepackage{lipsum}
\usepackage[percent]{overpic}
\usepackage[separate-uncertainty=true]{siunitx}
\usepackage{bibunits}
\usetikzlibrary{matrix}
\newcommand\myshade{85}
\colorlet{mylinkcolor}{violet}
\colorlet{mycitecolor}{YellowOrange}
\colorlet{myurlcolor}{Aquamarine}
\hypersetup{
  linkcolor  = mylinkcolor!\myshade!black,
  citecolor  = mycitecolor!\myshade!black,
  urlcolor   = myurlcolor!\myshade!black,
  colorlinks = true,
}

\DeclareMathAlphabet\mathbfcal{OMS}{cmsy}{b}{n}

\makeatletter

\AtBeginDocument{}
\RS@ifundefined{subsecref}
  {\newref{subsec}{name = \RSsectxt}}
  {}
\RS@ifundefined{thmref}
  {\def\RSthmtxt{theorem~}\newref{thm}{name = \RSthmtxt}}
  {}
\RS@ifundefined{lemref}
  {\def\RSlemtxt{lemma~}\newref{lem}{name = \RSlemtxt}}
  {}

\makeatother
\newcommand{\appx}{Supplementary Material}

\newcommand{\ptpt}[1]{\left( #1 \right)}
\newcommand{\pqpq}[1]{\left[ #1 \right]}

\newcommand{\fraparsq}[2]{\dfrac{\partial ^2 #1}{\partial #2^2}}

\newcommand{\fraparcu}[2]{\dfrac{\partial ^3 #1}{\partial #2^3}}

\begin{document}
\title{Interrelation of elasticity and thermal bath in nanotube cantilevers }
\author{S. Tepsic}
\affiliation{ICFO - Institut De Ciencies Fotoniques, The Barcelona Institute of Science and Technology, 08860 Castelldefels (Barcelona), Spain}
\author{G. Gruber}
\author{C. B. M\o ller}
\affiliation{ICFO - Institut De Ciencies Fotoniques, The Barcelona Institute of Science and Technology, 08860 Castelldefels (Barcelona), Spain}
\author{C. Mag\'en}
\affiliation{Instituto de Nanociencia y Materiales de Arag\'on (INMA), CSIC-Universidad de Zaragoza, 50009 Zaragoza, Spain}
\affiliation{Laboratorio de Microscop\'ias Avanzadas (LMA), Universidad de Zaragoza, 50018 Zaragoza, Spain}
\author{P. Belardinelli}
\affiliation{DICEA, Polytechnic University of Marche, 60131 Ancona, Italy}
\author{E. R. Hern\'{a}ndez}
\affiliation{Instituto de Ciencia de Materiales de Madrid (ICMM-CSIC), 28049 Madrid, Spain}
\author{F. Alijani}
\affiliation{Department of Precision and Microsystems Engineering, 3ME, Mekelweg 2, (2628 CD) Delft, The Netherlands}
\author{P. Verlot}
\affiliation{School of Physics and Astronomy - The University of Nottingham, University Park, Nottingham NG7 2RD, United Kingdom}
\author{A. Bachtold}
\affiliation{ICFO - Institut De Ciencies Fotoniques, The Barcelona Institute of Science and Technology, 08860 Castelldefels (Barcelona), Spain}

\begin{abstract}
We report the first study on the thermal behaviour of the stiffness of individual carbon nanotubes, which is achieved by measuring the resonance frequency of their fundamental mechanical bending modes. We observe a reduction of the Young's modulus over a large temperature range with a slope $-(173\pm 65)$~ppm/K in its relative shift. These findings are reproduced by two different theoretical models based on the thermal dynamics of the lattice. These results reveal how the measured fundamental bending modes depend on the phonons in the nanotube via the Young's modulus. An alternative description based on the coupling between the measured mechanical modes and the phonon thermal bath in the Akhiezer limit is discussed.          
\end{abstract}
\maketitle

\begin{bibunit}[apsrev4-2]

In engineering, thermoelasticity is central in determining the elastic limits of structures ranging from large scale spacecrafts \cite{thornton1996thermal} and nuclear plants \cite{zudans1965thermal} down to nano-structured systems. A rich underlying phenomenology emerges for small structures, including dissipation \cite{Zener1938,Lifshitz2000}, fluctuations \cite{Kubo1966,Cleland2002}, and torque generation \cite{FANG1999,Murozono1994}, which are key to the development of state-of-the-art nano- and micro-electromechanical technologies \cite{Saulson1990,Albrecht1991}. Thermoelasticity has also been used with success in condensed matter physics, where thermal measurements of the stiffness unveil the phase transition of charge-density waves and superconductivity in transition metal dichalcogenides and high-$T_c$ superconductors \cite{Barmatz1975,Brill1984,Hoen1988}.     
From a fundamental point of view, the thermal behaviour of the stiffness -- quantified by the Young's modulus -- emerges from the non-trivial interplay of the binding energy and the lattice dynamics. However, the effect of the thermal lattice dynamics on the stiffness has remained elusive in individual nanoscale systems due to experimental challenges related to manipulating and measuring such small objects. 

In this work, we use the exquisite sensing capabilities of mechanical resonators based on nanoscale systems \cite{Chiu2008,Gil-Santos2010,Chaste2012,Moser2013,Yeo2013,Siria2012,Cole2015,Weber2016,Lepinay2016,Rossi2016,Dolleman2017,Bonis2018,Blaikie2019,Rossi2019,Sahafi2020,fogliano2020} to resolve the small effect associated with the thermal behaviour of their stiffness. Using the resonance frequency measured by optomechanical spectroscopy, we estimate the Young's modulus of micrometer-long nanotube cantilevers from room temperature down to a few Kelvins. These results agree with the temperature dependence of the resonance frequency predicted by molecular dynamics simulations, which take into account the lattice dynamics of the nanotube. Our measurements are also consistent with the Young's modulus directly computed from a quasi-harmonic approximation of the free energy of the phonon modes. This work not only shows how the stiffness of an individual nanotube is related to its phonons, but it also highlights the role of the phonon thermal bath in nanotube cantilevers, which is a topic of importance in the field of nanomechanical resonators \cite{Chiu2008,Gil-Santos2010,Chaste2012,Moser2013,Yeo2013,Siria2012,Cole2015,Weber2016,Lepinay2016,Rossi2016,Dolleman2017,Bonis2018,Blaikie2019,Rossi2019,Sahafi2020,fogliano2020}.

\begin{figure}[tb]
	\begin{center}
\includegraphics[width=.9\linewidth]{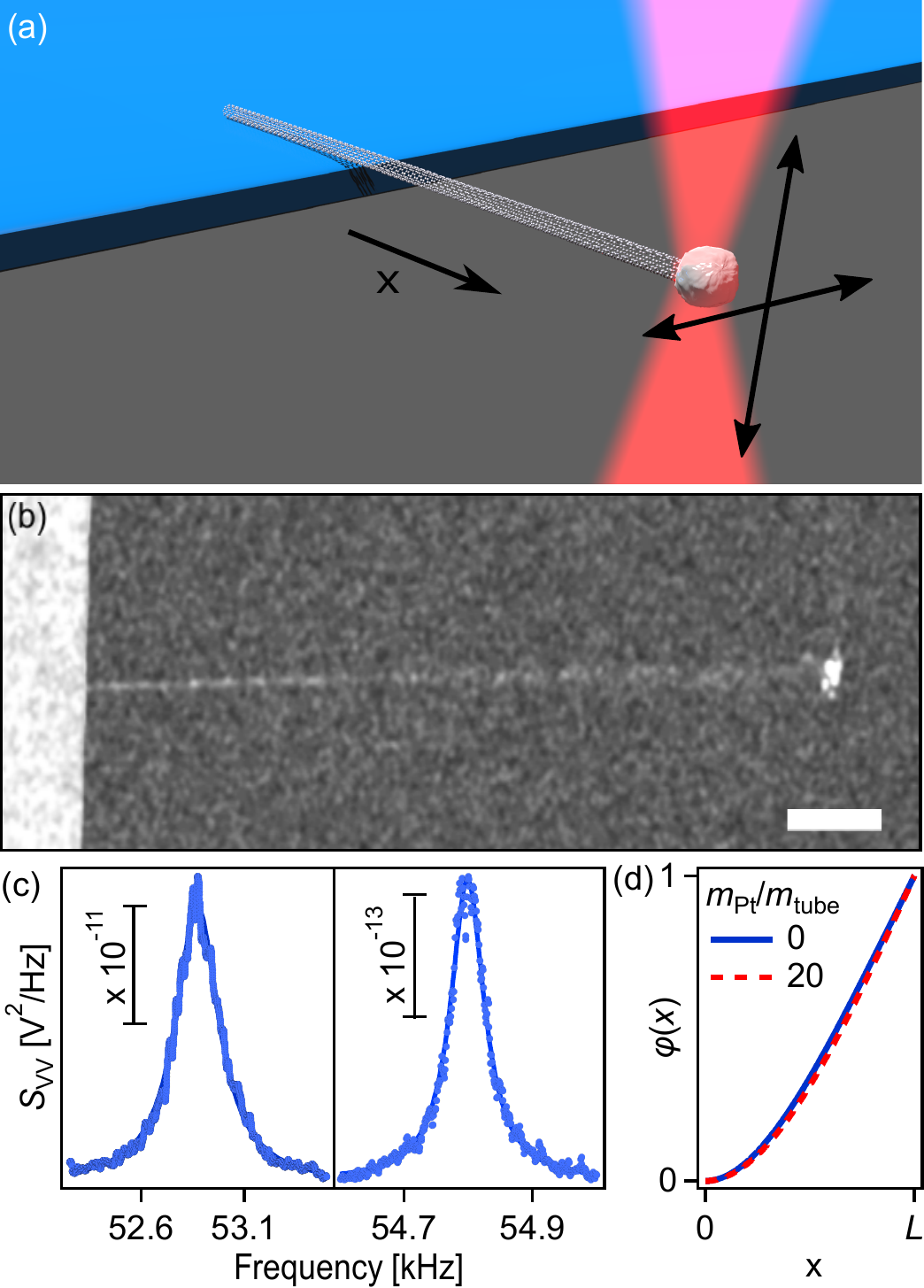}
\caption{(a) Schematic of the experimental setup. The sample is placed at the waist of a strongly focused beam of a He-Ne laser. The scattered light (not shown) is collected in reflection by means of an optical circulator and further sent on an avalanche photodetector. The two double arrows represent the polarization of the fundamental mode doublet. (b) Device A imaged by scanning electron microscopy after the deposition of a platinum nanoparticle; the scale bar is \SI{1}{\micro\meter} \cite{Gruber2019}. (c) Power spectra of the optical reflection from device A showing the resonance of the thermal motion of the fundamental mode doublet at \SI{300}{K}. The two spectra are recorded using different positions of the nanotube in the laser waist to enhance the signal \cite{Tavernarakis2018}. (d) Calculated profile $\varphi (x)$ of the fundamental mode shape along the nanotube axis estimated for two different platinum particle masses normalized by the nanotube mass.}
\label{fig:setup_layout}
\label{fig:Mode}
	\end{center}
\end{figure}

We use the single clamped resonator layout, where one end of the nanotube is attached to a silicon chip and the other end is free. This layout avoids prestress in the nanotube built-in during fabrication, in contrast to what may happen with the double clamped layout. As a result, the restoring force is given solely by the bending rigidity. This enables us to probe the Young’s modulus $Y$ by measuring the resonance frequency, $\omega_0 \propto \sqrt{Y}$ \cite{Lifshitz2008}. Such a resonance-based methodology is also employed in thermoelasticity studies on larger scale systems \cite{Barmatz1975,Brill1984,Hoen1988,Wachtman1961}.  

We engineer a platinum particle at the free end of the nanotube, so that the resonator can be measured by scattering optomechanical spectroscopy (Fig.~\ref{fig:setup_layout}a) \cite{Tavernarakis2018}. We grow the particle by focused electron beam-induced deposition \cite{Gruber2019}. Figure~\ref{fig:setup_layout}b shows a scanning electron microscopy image of device~A. Transmission electron microscopy (TEM) indicates that nanotubes can be made from one to a few walls, with a median value of two walls (Supplementary Material, Sec.~I \cite{SM}). The vibrations are detected by measuring the backscattered intensity from a \SI{632}{nm} laser beam focused onto the particle. Figure~\ref{fig:setup_layout}c shows the optomechanical spectrum of device~A. The resonance frequencies of the fundamental modes polarized in perpendicular directions are about \SI{52.9}{\kilo\hertz} and \SI{54.8}{\kilo\hertz}. The platinum particle does not affect the restoring force nor the eigenmode shape of the two fundamental modes (Fig.~\ref{fig:setup_layout}d), in contrast to what happens for higher frequency modes (Supplementary Material, Sec.~II \cite{SM}). In this work we use low laser power so that the resonance frequency is not affected by absorption heating and optical backaction  \cite{Tavernarakis2018}.\nocite{Tsioutsios2017, balachandran2008vibrations, ramanAPL2007, Wang2010, Yang2011, Tavernarakis2014, Atalaya2011, plimpton2007lammps, TersoffPRB1988, LindsayPRB2010, TangPRB2009, YakobsonPRL1996, Klessig, doi:10.1063/1.449071, sajadi2019nonlinear, dftb, phon, Vacher2005} 

We quantify $Y=\SI{1.06(28)}{\tera\pascal}$ at room temperature from six devices by combining thermal motion variance measurement and TEM imaging; the advantage of this method is that it does not rely on the cantilever mass (Supplementary Material, Sec.~III \cite{SM}). The estimated Young's modulus is similar to previously predicted and measured values \cite{Treacy1996, Krishnan1998, Salvetat1999,Lu1997,Hernndez1998,SnchezPortal1999}. This indicates that the contamination adsorbed on the nanotube surface has little contribution to the stiffness of the nanotube. The contamination, which is localized along some portions of nanotubes as observed by TEM, presumably consists of hydrocarbons adsorbed during their exposure to air and the particle growth. The typical stiffness reported for such amorphous material is comparatively low $Y\approx\SIrange[range-units=brackets, range-phrase=\text{ -- }]{50}{300}{\giga\pascal}$~\cite{Robertson2002}.

\begin{figure}[tb]
	\begin{center}
\includegraphics[width=.9\linewidth]{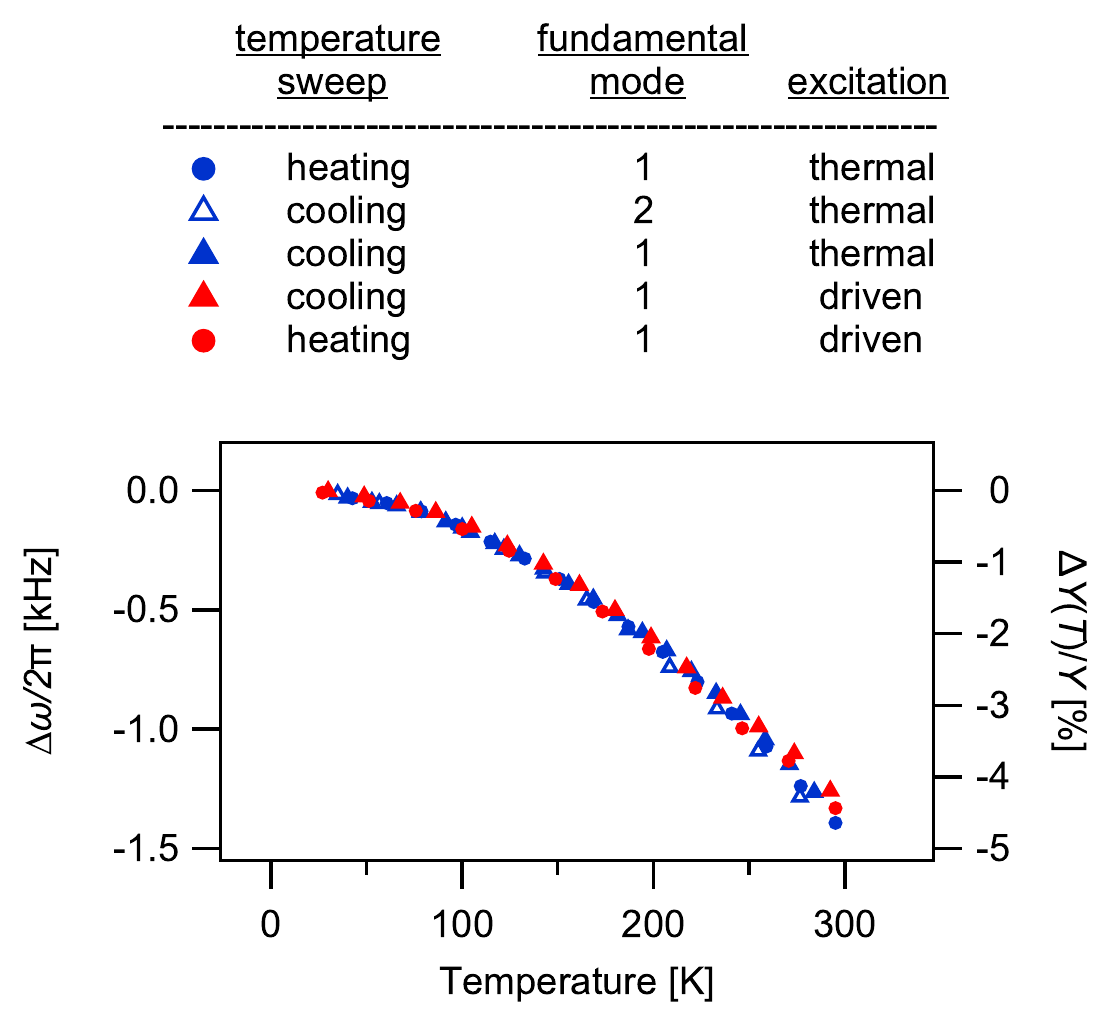}
\caption{Resonance frequency and relative change of the Young's modulus of device A as a function of cryostat temperature. The legend indicates the direction of the temperature sweep (cooling or heating), which fundamental mode is measured, and whether the detected vibrations are thermal or driven with a piezo-actuator.}
\label{fig:main}
	\end{center}
\end{figure}

\begin{figure*}[bt]
	\begin{center}
\includegraphics[width=.8\textwidth]{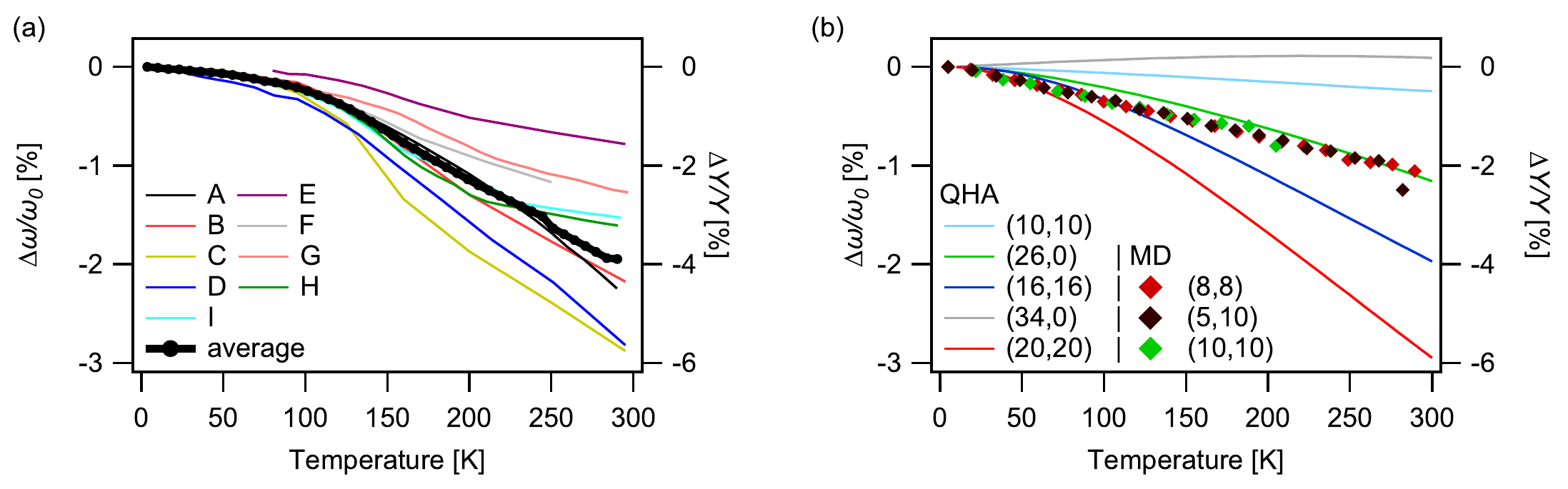}
\caption{Comparison of the relative change of the resonance frequency and the Young's modulus between experiment (a) and theory (b) for different nanotubes. The theoretical results are obtained for different nanotube chiralities with either molecular dynamics (MD) simulations or quasi-harmonic approximation (QHA) calculations. The MD simulations and the QHA calculations quantify $\Delta \omega_0/\omega_0$ and $\Delta Y/Y$, respectively. }
\label{fig:DeltaELin}
	\end{center}
\end{figure*}

Figure~\ref{fig:main} shows the variation of the resonance frequency of device~A when sweeping the temperature $T$. The variation is remarkably similar for both fundamental modes, independent of the temperature sweep direction and of whether the motion is thermal or driven with a piezo-actuator. This variation of the resonance frequency $\omega_0=\sqrt{k/m}$ is associated to the change of the spring constant $k$, which is linearly proportional to $Y$ in the single clamped layout. We extract the relative shift of the Young's modulus from the relation $\frac{\Delta Y(T)}{Y(T_{min})}=2\frac{\Delta\omega_0(T)}{\omega_0(T_{min})}$, where $T_{min}$ is the lowest temperature at which we record the vibrations. Figure~\ref{fig:DeltaELin}a shows the measurements of nine different devices. They all feature the same trend with a reduction of the Young's modulus when increasing temperature. The dependence is essentially linear above about \SI{100}{K}; the slope averaged over devices is $\Delta Y(T)/Y\cdot1/T=-(173\pm 65)$~ppm/K. These measurements are related to neither the mass adsorbed on the nanotube nor the diffusion of adsorbed atoms along the nanotube nor the thermal expansion of the nanotube nor the combination of the Duffing nonlinearity and the thermal motion, as shown in Sec.~IV of Supplementary Material \cite{SM}.

These measurements can be captured by molecular dynamics~(MD) simulations of the nanotube cantilever dynamics. The temperature dependence of the resonance frequencies of the lowest energy bending modes obtained from the MD simulations behave in the same way as those we measure (Figs.~\ref{fig:DeltaELin}a,b). The associated slope estimated for different nanotube chiralities leads to $\Delta Y(T)/Y\cdot1/T=-(79\pm 6)$~ppm/K, which is rather similar to the measured value. This suggests that the thermal behaviour of the Young's modulus in our measurements is related to the lattice dynamics of nanotubes.

We employ a second method to directly compute the Young's modulus from the energy dispersion of the nanotube phonon modes. For this, we evaluate the free energy $F(T,\epsilon)$ of the phonon modes at $T$ and strain~$\epsilon$ with the quasi-harmonic approximation, yielding 
\begin{equation*}
Y(T) = \frac{1}{V_0(T)} \left( \frac{\partial^2 F(T,\epsilon)}{\partial \epsilon^2} \right)_{\epsilon=0},
\end{equation*}
where $V_0(T)$ is the equilibrium volume at this temperature. The resulting $Y(T)$ dependence is also consistent with the measurements (Figs.~\ref{fig:DeltaELin}a,b). The slope for different chiralities is $\Delta Y(T)/Y\cdot1/T=-(104\pm 102)$~ppm/K. The variation of the slope is larger than that obtained with molecular dynamics; this difference may be due to the infinite nanotube length and the purely linear vibrational dynamics considered in the quasi-harmonic approximation method, while the lengths in the molecular dynamics simulations are much shorter, that is, less than \SI{40}{nm}. Overall, the experimental findings are fairly consistent with both models considering the typical differences between the values of $Y$ of nanotubes obtained with different experimental and theoretical methods \cite{Treacy1996, Krishnan1998, Salvetat1999,Lu1997,Hernndez1998,SnchezPortal1999}.  Both theoretical models are described in the Supplementary Material (Secs.~V and VI) \cite{SM}.

\begin{figure}[htb]
	\begin{center}
\includegraphics[width=.9\linewidth]{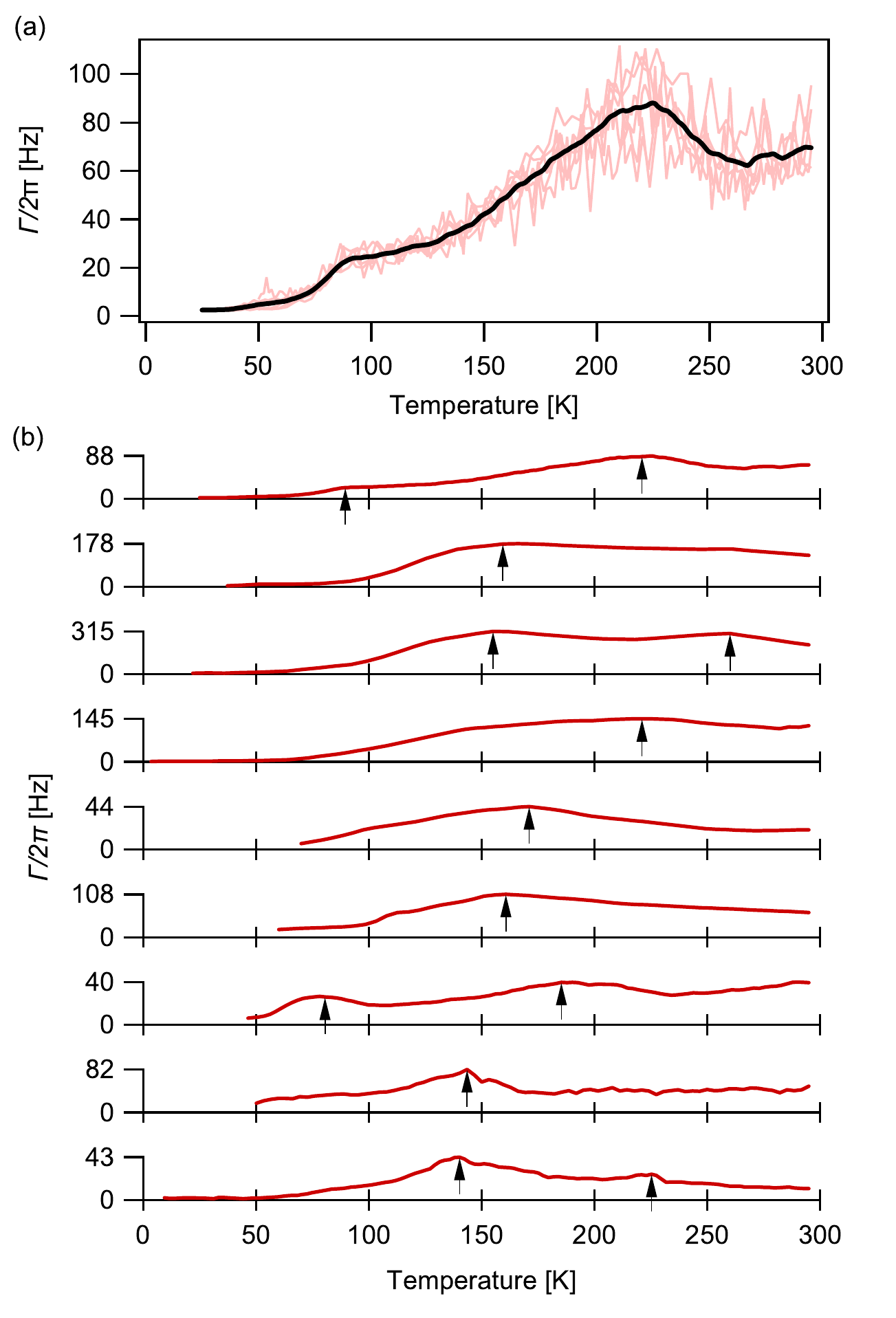}
\caption{(a) Temperature dependence of the mechanical linewidth for device A. The black line is the average of different temperature traces (red lines). (b) Temperature dependence of the linewidth for all the measured nanotubes, from device~A at the top to device~I at the bottom; the associated resonance frequencies are \SI{54}{\kilo\hertz}, \SI{96}{\kilo\hertz}, \SI{194}{\kilo\hertz}, \SI{77}{\kilo\hertz}, \SI{57}{\kilo\hertz}, \SI{108}{\kilo\hertz}, 
\SI{44}{\kilo\hertz}, 
\SI{58}{\kilo\hertz},
\SI{48}{\kilo\hertz}. The arrows indicate peaks in dissipation.}
\label{fig:fwhm}
	\end{center}
\end{figure}

These results show how the measured fundamental mechanical modes are linked to phonons via the Young's modulus. An alternative way to describe this link is to consider the coupling of the measured mechanical modes with the thermal bath made of the phonons of the nanotube. In other words, the measured $T$ dependence of $\omega_0$ is related to the phonon thermal bath. It is likely that the phonon thermal bath in our experiments operate in the Akhiezer limit \cite{akhiezer1938absorption}. Over the temperature range that we measure, the phonon modes in nanotubes with energy $\hbar\omega_{k}$ similar to $k_BT$ have decay rates $1/\tau_k$  larger than $\omega_{0}$, since $\tau_k \approx 10$~ns was measured for breathing modes at $T=5$~K \cite{LeRoy2004} and we estimate $\tau_k$ to be typically in the $10-1000$~ns range for the longitudinal and twist modes \cite{demartino2009} (Sec.~VII of Supplementary Material \cite{SM}). (The estimation of $\tau_k$ for high-energy bending modes is complicated and beyond the scope of this work.) This sets the Akhiezer limit $\omega_{0} \tau_k \ll 1$ at least for the breathing, longitudinal, and twist modes \cite{Atalaya2016}. It involves three-phonon processes, where one vibration quantum of the measured mode is absorbed together with the absorption and the emission of high-energy phonons with frequencies $\omega_{k}$ and $\omega_{k^{\prime}}$, respectively. The sizeable decay rates of the high-energy phonons lead to uncertainty in their energy. This lifts to some extent the restriction associated with the energy conversation of the three-phonon process, $\omega_{0}= \omega_{k}-\omega_{k^{\prime}}$, which holds in the Landau-Rumer limit when $\omega_{0} \tau_k \gg 1$. For this reason, the resonance frequency reduction and the relaxation in the Akhiezer limit are expected to be larger than that in the Landau-Rumer limit over the studied temperature range. The thermoelastic limit \cite{Lifshitz2000} does not apply for nanotubes, since the model relies on phonons that locally reach thermal equilibrium at different temperatures on the two sides of the beam cross-section, which is not realistic for such narrow resonators. 

It is expected that the phonon thermal bath significantly contributes to the measured dissipation via the Akhiezer relaxation, since a thermal bath results in a resonance frequency reduction as well as dissipation, both of them being related through the Kramers–Kronig relations \cite{Landau1986}. Figures~\ref{fig:fwhm}a,b show the measured temperature dependence of the mechanical linewidth of the different measured devices. The measurements feature one or two peaks of dissipation at some specific temperatures. These observed peaks could arise from the Akhiezer relaxation. The Akhiezer dissipation rate depends in a complicated way on the number of phonon modes with energy $\hbar\omega_{k} \lesssim k_BT$, their population, and their decay rate \cite{Atalaya2016}. The temperature dependence of the Akhiezer dissipation rate could feature one or more peaks in dissipation, especially since the phonon density of states varies up and down as a function of energy \cite{ando2002,demartino2009} and the temperature behaviour of the decay rate changes for different phonon modes. In addition, the dissipation peaks could emerge at different temperatures for different nanotube chiralities, since the phonon energy dispersion is chirality dependent. The measured peaks in dissipation cannot be described by the model that is used in the literature \cite{Faust2014,Hamoumi2018} to quantify dissipation due to defects. See Secs.~VIII and IX of Supplementary Material for further discussion on the Akhiezer dissipation and dissipation due to defects \cite{SM}.

In conclusion, we report the first experimental study of the temperature dependence of the Young's modulus of a nanoscale system. The measurements are consistent with theoretical predictions based on the nanotube lattice dynamics. This indicates that the phonon thermal bath plays an important role in the dynamics of nanotube cantilevers, including thermal vibrational noise, dissipation, and resonance frequency reduction. Further theoretical work is needed to compute the Akhiezer relaxation in nanotubes beyond the models used so far, where a single decay rate is employed for all the high-frequency phonon modes \cite{Hamoumi2018,Rodriguez2019,Iyer2016}. This may be achieved with a microscopic theory \cite{Atalaya2016} taking into account the phonon energy dispersion \cite{ando2002} and the energy decay of high-frequency phonons \cite{demartino2009}. It will be interesting to see whether such a model leads to dissipation peaks at specific temperatures as observed in our work.  

We thank Mark Dykman and Andrew Fefferman for enlightening discussions. This work is supported by ERC advanced (grant number 692876), ERC PoC (grant number 862149), Marie Skłodowska-Curie PROBIST (grant number 754510), the Cellex Foundation, the CERCA Programme, AGAUR (grant number 2017SGR1664), Severo Ochoa (grant number SEV-2015-0522), MICINN (grant number RTI2018-097953-B-I00 and PGC2018-096955-B-C44), the Fondo Europeo de Desarrollo Regional, the grant MAT2017-82970-C2-2-R of Spanish MINECO and the project E13\textunderscore17R from Aragon Regional Government (Construyendo Europa desde Arag\'{o}n). F.A. acknowledges support from European Research Council (ERC) starting grant number 802093.

\end{bibunit}

\clearpage

\begin{widetext}

\begin{bibunit}[apsrev4-2]

\begin{Large}
\appx
\end{Large}

\section{Device fabrication and structural characterization}
\label{sec:Fabrication}

The carbon nanotubes were grown on silicon substrates via chemical vapor deposition. A Zeiss Auriga scanning electron microscope (SEM) was used to select suitable nanotube cantilevers. The SEM is equipped with a gas injection system, which was used to deposit platinum particles at the apex of the nanotubes for their optomechanical functionalization \cite{SMTavernarakis2018, SMGruber2019}. Figure \ref{fig:SEM}(a) shows a pristine nanotube cantilever (device A). Figure \ref{fig:SEM}(b) shows the same cantilever after the deposition of a Pt particle. The free end of the cantilever is blurred in the SEM images due to the thermally driven motion. The displacement profile was measured by a SEM line trace across the nanotube at the tip. Figure \ref{fig:SEM}(c) shows the observed Gaussian distribution in the secondary electron current $I_\textrm{SE}$, as expected for thermal vibrations \cite{SMTsioutsios2017}. The displacement variance $\sigma^2=(\SI{87.4}{\nano\meter})^2$ was obtained from a fit of the data. The spring constant $k=\SI{5.42d-7}{\newton/\meter}$ was determined from the equipartition theorem $k={k_\textrm{B}T}/{\sigma^2}$ where $k_\textrm{B}$ is the Boltzmann constant and $T$ is the temperature \cite{SMTsioutsios2017}. The mass of the deposited Pt particle was controlled during its growth by monitoring the mechanical resonance frequency of the lowest flexural mode of the nanotube; the thermal vibrations were measured by pointing the electron
beam onto the apex of the nanotube in spot mode while
recording the noise of $I_\textrm{SE}$~ \cite{SMGruber2019}. The initial effective mass of the nanotube was $m_0^*=\SI{243}{\atto\gram}$ and the mass of the deposited particle visible in Fig. \ref{fig:SEM}(b) was $m_\textrm{Pt}=\SI{3.6\pm1.1}{\femto\gram}$. All discussed samples were fabricated as described above. The mechanical properties of the samples that were optomechanically characterized at low temperature (devices A-I) are summarized in Table \ref{tab:Devices}.

\begin{figure*}[b]
\begin{center}
\includegraphics[width=0.7\textwidth]{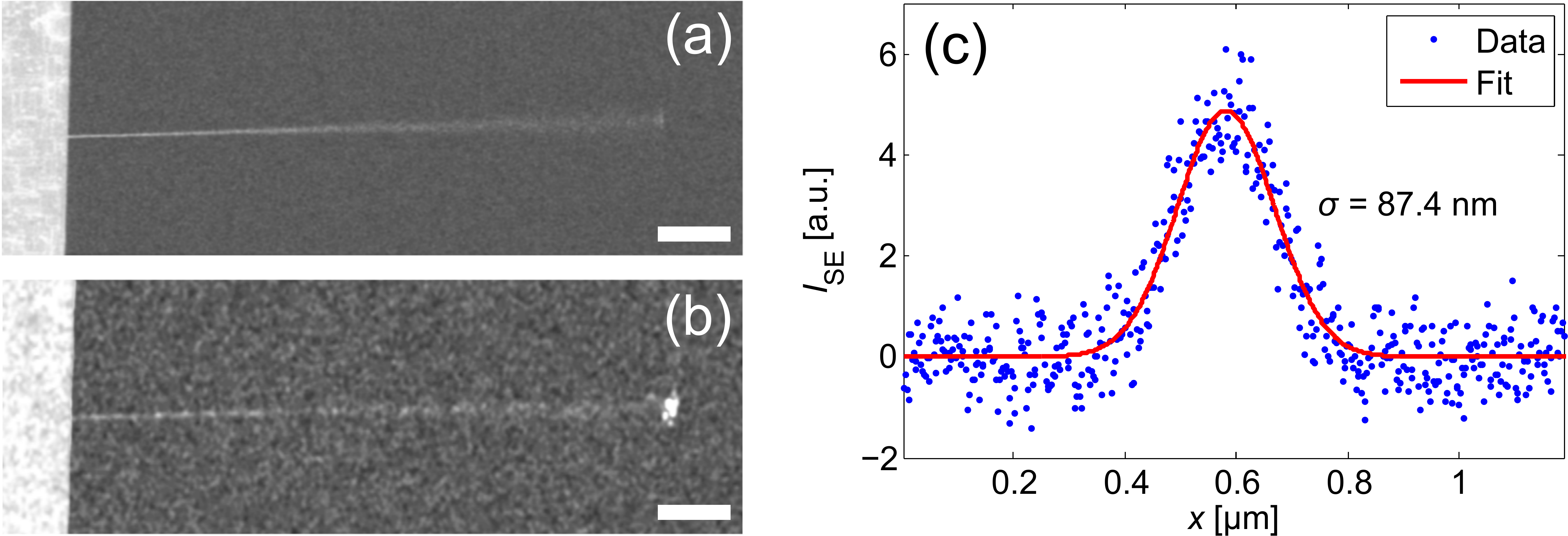}
\caption{Device A imaged by SEM (a) before and (b) after deposition of a Pt nanoparticle; the scale bars are \SI{1}{\micro\meter}. (c) Secondary electron signal $I_\textrm{SE}$ across the apex of the nanotube;  from a Gaussian fit the displacement variance $\sigma^2=(\SI{87.4}{\nano\meter})^2$ is determined. }
\label{fig:SEM}
\end{center}
\end{figure*}

\begin{table}[b]
\caption{\label{tab:Devices} Mechanical properties of the nanotube cantilever devices discussed in the main text. These include the length $l$, the standard deviation of the thermal displacement $\sigma$, the spring constant $k$, and the mass ratio $m^*$ between the Pt particle and the nanotube as defined in Supplementary Section \ref{sec:eigenmode}.}
\begin{ruledtabular}
\begin{tabular}{lrrrr}
Device &  $l\ (\mu$m) &$\sigma$ (nm) &$k$ (N/m) & $m^*$ \\
\hline
A & 8.2 &  87.4 & $5.42\times10^{-7}$  & 3.7 \\%(Tube3) 
B & 6.5 &  22.8 & $7.92\times10^{-6}$  & 4.6 \\%(SL4) 
C & 2.4 &  16.0 & $1.64\times10^{-5}$  & 83.0 \\%(H1B)
D & 7.8 &  55.8 & $1.33\times10^{-6}$  & 5.5 \\%(AFM6)
E & 10.0 &  44.2 & $2.10\times10^{-6}$ & 4.6 \\%(Tube16)
F & 5.0 &  34.9 & $3.39\times10^{-6}$  & 8.8 \\%(Tube19) 
G & 5.2 &  80.1 & $6.46\times10^{-7}$  & 18.3 \\%(Tube5) 
H & 4.6 &  73.5 & $7.63\times10^{-7}$  & 60.9 \\%(Tube6) 
I & 11.9 &  90.0 & $5.11\times10^{-7}$  & 3.7 \\%(Tube24) 
\end{tabular}
\end{ruledtabular}
\end{table}

We performed high-resolution transmission electron microscopy (HRTEM) to assess the microscopic structure of nanotube cantilevers. The samples were fabricated on silicon aperture windows using the identical procedure as outlined above. HRTEM imaging was conducted using a Thermo Fisher Titan Cube 60-300, equipped with an image aberration corrector CETCOR from CEOS. The microscope was operated at $\SI{80}{\kilo\volt}$ to minimize beam damage and achieve a spatial resolution below $\SI{1.4}{\angstrom}$ . Figure \ref{fig:TEM} shows atomically resolved images obtained for different devices near the clamping point where the thermal displacement is negligible. The devices shown are a single wall device, a seven wall device, and a triple wall device. The latter device was also characterized optomechanically at low temperature before conducting the HRTEM experiments and is referred to as device C in the main text and in table \ref{tab:Devices}. The amorphous material visible in Figs. \ref{fig:TEM} (a) and (c) presumably consists of hydrocarbons adsorbed during their exposure to air and the particle growth~\cite{SMGruber2019}.

Using such HRTEM images, we determined the number of walls and the associated diameters for six different devices ranging from single wall to seven wall nanotubes. Table \ref{tab:TEM} shows the diameters obtained by HRTEM together with other parameters obtained by SEM. The calculation of the Young's modulus in the table is outlined in Section~\ref{sec:absY}.

\begin{figure}[hb]
\begin{center}
\includegraphics[width=0.7\linewidth]{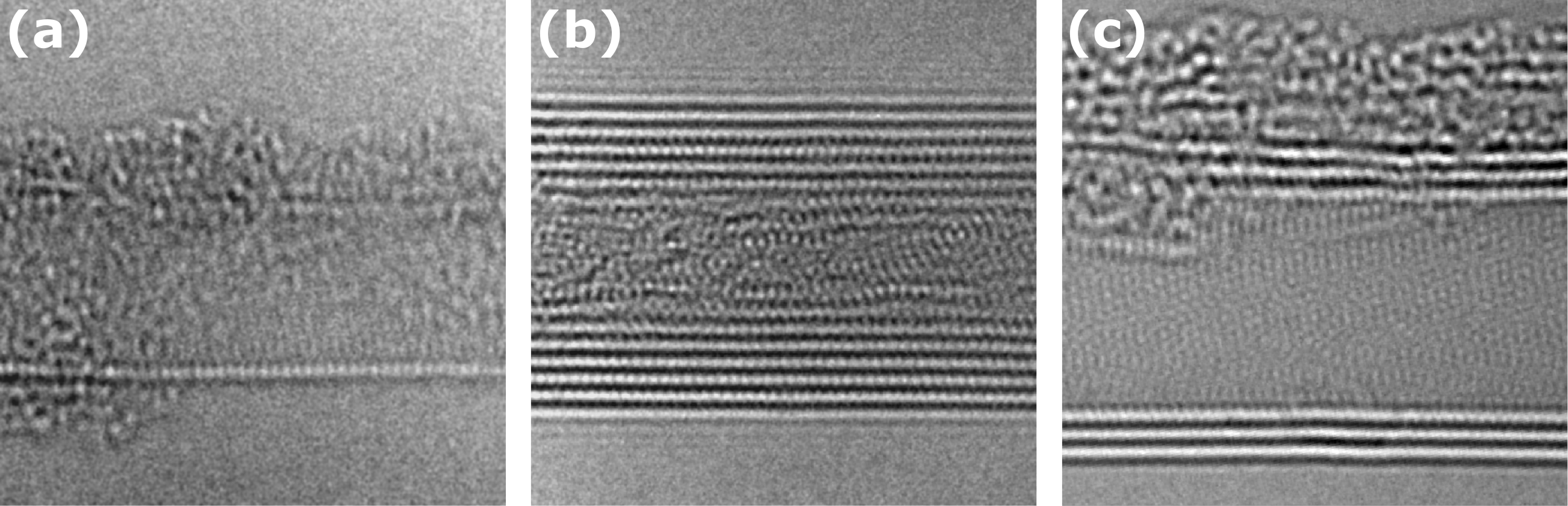}
\caption{HRTEM images recorded near the clamping point of single wall device T1 (a), seven wall device T6 (b) and triple wall device C (c). In order to enhance the signal-to-noise ratio multiple images were overlaid and averaged. All image dimensions are $\SI{10}{\nano\meter}$ by $\SI{10}{\nano\meter}$.}
\label{fig:TEM}
\end{center}
\end{figure}

\begin{table}[hb]
\caption{\label{tab:TEM} Properties of different nanotube cantilevers functionalized with platinum particles. Cantilever length $l$ and spring constant $k$ were obtained by SEM imaging. Number of walls $N$ and associated diameters $d_i$ were obtained by HRTEM. The Young's modulus $Y$ was calculated as described in section \ref{sec:absY}.}
\begin{ruledtabular}
\begin{tabular}{lrrrrrrr}
Device & $N$ & $l\ (\mu$m) &$k$ (N/m) &  $d_i$ (nm) & $Y$ (TPa)\\
\hline
T1 & 1 & $2.8\pm0.1$ &  $(6.51\pm1.53)\times10^{-7}$  &  $d_1=3.24\pm0.14$  & $1.04\pm0.49$ \\%(Tube13_2)
\hline
T2 & 2 & $2.2\pm0.1$ &  $(1.93\pm0.42)\times10^{-6}$ &  $d_1=3.37\pm0.16$, $d_2=2.58\pm0.16$ & $0.91\pm0.46$ \\%(Tube13_3) 
\hline 
T3 & 2 & $1.8\pm0.1$ &  $(3.33\pm0.52)\times10^{-6}$  &  $d_1=3.58\pm0.11$, $d_2=2.80\pm0.09$  & $0.79\pm0.34$ \\%(Tube13_1) 
\hline 
T4 & 2 & $3.8\pm0.1$ &   $(7.42\pm1.73)\times10^{-7}$  &  $d_1=3.75\pm0.19$, $d_2=2.89\pm0.22$ & $1.31\pm0.64$ \\%(Tube14_1) 
\hline 

C & 3 & $2.4\pm0.1$ &  $(1.64\pm0.38)\times10^{-5}$  &  $d_1=5.84\pm0.09$, $d_2=5.15\pm0.08$, $d_3=4.38\pm0.11$ & $1.35\pm0.55$\\%(H1B)
\hline

T6 & 7 & $7.1\pm0.1$ &  $(7.36\pm1.83)\times10^{-7}$  &  $d_1=6.24\pm0.07$, $d_2=5.55\pm0.07$, $d_3=4.88\pm0.09$, & $0.96\pm0.33$\\

 & & & & $d_4=4.20\pm0.10$, $d_5=3.53\pm0.09$, $d_6=2.88\pm0.09$, \\
 & & & & $d_7=2.15\pm0.14$ \\%(Tube14_2) 
\end{tabular}
\end{ruledtabular}
\end{table}

\section{Eigenmodes and  spring constant of a cantilever with added mass at the free end}
\label{sec:eigenmode}

%In presence of a mass particle at the free end, the boundary condition $\fraparcu{y}{x}\mid_{x=L}=0$ changes to $- Y I\fraparcu{y}{x}\mid_{L=0}=m_{bead} c^2 {\alpha_n}^4 y_{x=L}$.
\subsection{Model}
The Euler-Bernoulli partial differential equation (PDE) that describes the motion $y\ptpt{x,t}$ of a vibrating beam is
\begin{equation}
\fraparsq{y}{t}+\dfrac{Y I}{\rho A}\dfrac{\partial^4 y}{\partial x^4}=0.
\label{EB1}
\end{equation}
In Eq.~\ref{EB1}, $Y$ is the Young’s modulus, $I$ is the second moment of the cross-sectional area $A$, and $\rho$ is the density of the carbon nanotube (CNT) with length $l$. Solution of Eq.~\ref{EB1} is  
\begin{equation}
y\ptpt{x,t}=\cos \omega_n t \pqpq{c_1 \cos \alpha_n x+c_2 \sin \alpha_n x+c_3 \cosh \alpha_n x+c_4 \sinh \alpha_n x}, 
\label{solEB1}
\end{equation}
with radial frequency $\omega_n=c {\alpha_n}^2$ and $c=\sqrt{\dfrac{Y I}{\rho A}}$. In Eq. \ref{solEB1}, $\alpha_n$ is the wave number whereas  $c_1, .. c_4$ are constants that will be determined by satisfying boundary conditions. In the presence of a particle with mass $m_{bead}$ at the free end, the boundary conditions to satisfy become: $y\mid_{x=0}=\partial {y}/\partial {x} \mid_{x=0}=0$, and $\partial^2 {y}/\partial {x^2} \mid_{x=l}=0$, $- Y I\fraparcu{y}{x}\mid_{x=l}=m_{bead} c^2 {\alpha_n}^4 y_{x=l}$ \cite{SMbalachandran2008vibrations}, in which the effect of the bead's rotary inertia  is neglected. Implementing these conditions in Eq. \ref{solEB1} leads to the following characteristic equation

\begin{equation}
\cos \Omega_n  \cosh \Omega_n +1+m^* \Omega_n \ptpt{\sinh \Omega_n \cos\Omega_n -\sin \Omega_n \cosh \Omega_n}=0 ,
\label{dispersion}
\end{equation}
where eigenvalues $\Omega_n=\alpha_n l$ are solutions of Eq.\ref{dispersion} with $m^*=m_{bead}/m_{beam}$. 
The eigenmodes associated with the eigevalues can then be obtained as 
\begin{equation}
\Phi_n(x)=\cos (\Omega_n x)-\cosh (\Omega_n x)-\frac{\cos (\Omega_n)+\cosh (\Omega_n) }{\sin (\Omega_n)+\sinh (\Omega_n)}(\sin (\Omega_n x)-\sinh (\Omega_n x)).
\label{eigenfrequencies}
\end{equation}
Figure~\ref{fig:eigenfrequencies} shows the variation of the first three eigenfrequencies as a function of $m^*$. When the ratio between the mass of the bead at the free end and the mass of the beam becomes large, the mode shapes approach those of a beam clamped at one end and hinged at the other. The mode shapes for an increasing $m^*$ are shown in Figs.~\ref{fig:eigenmodes}. The profile of the fundamental eigenmode is basically unchanged when increasing $m^*$, in contrast to what happens for the other eigenmodes.

\begin{figure}[b]
\begin{center}
\includegraphics[width=0.35\linewidth]{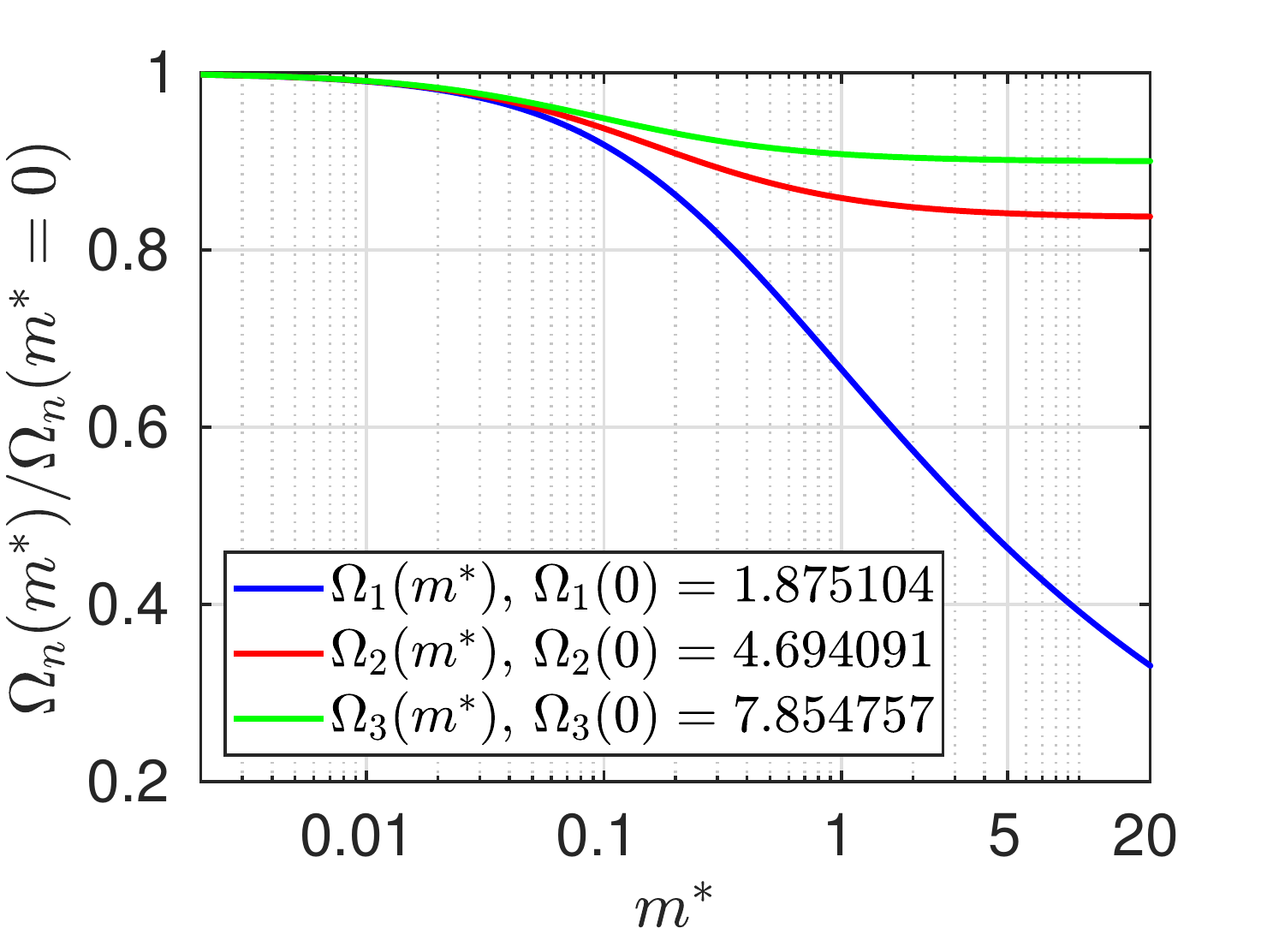}
\vspace{-0.5cm}
\caption{Influence of $m^*$ on the first three eigenfrequencies of a beam with added mass ($m_{bead}$) at the free end.}

\label{fig:eigenfrequencies}
\end{center}
\end{figure}

\begin{figure}[th]
\begin{center}
\includegraphics[width=0.32\linewidth]{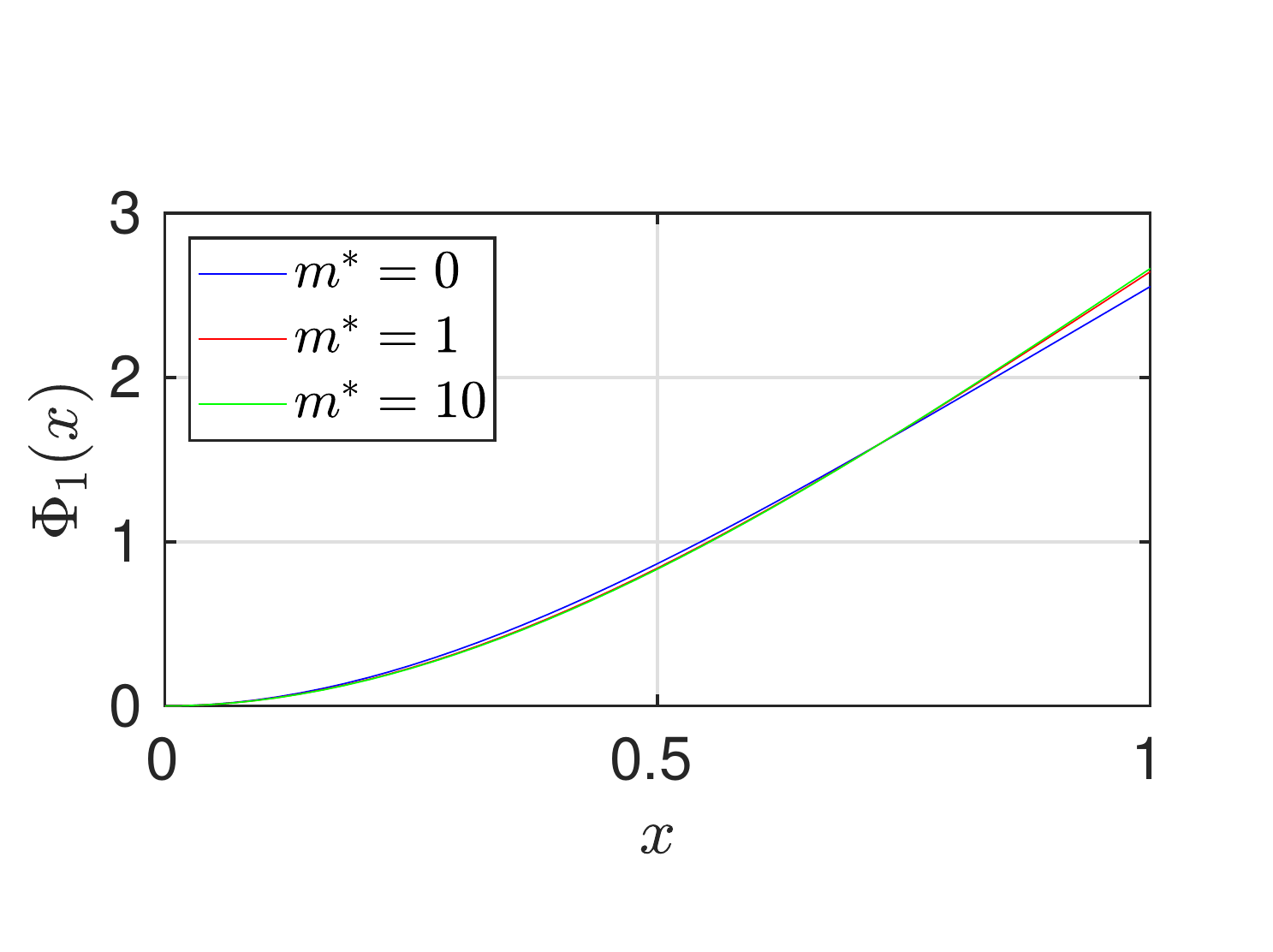}
\includegraphics[width=0.32\linewidth]{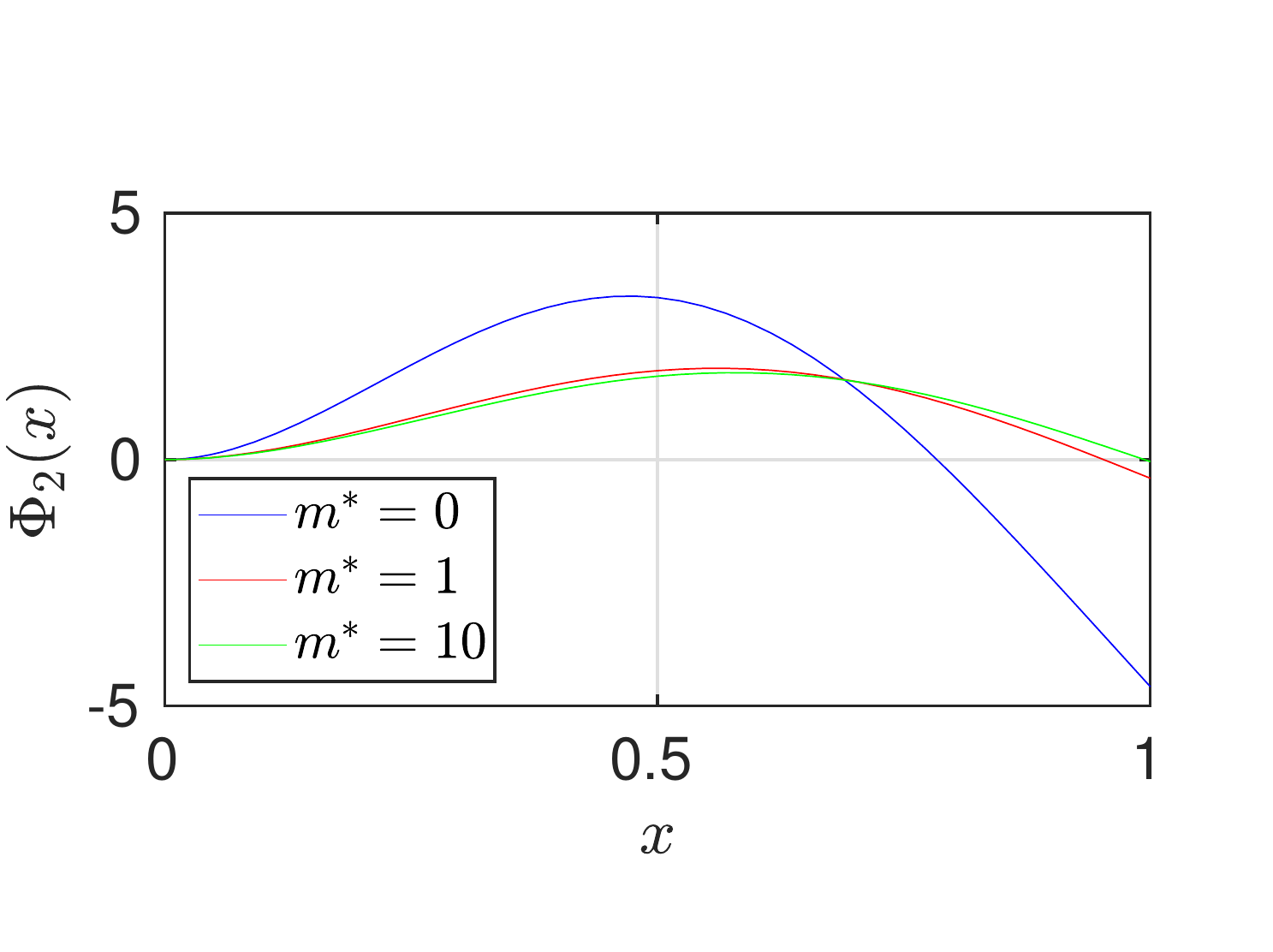}
\includegraphics[width=0.32\linewidth]{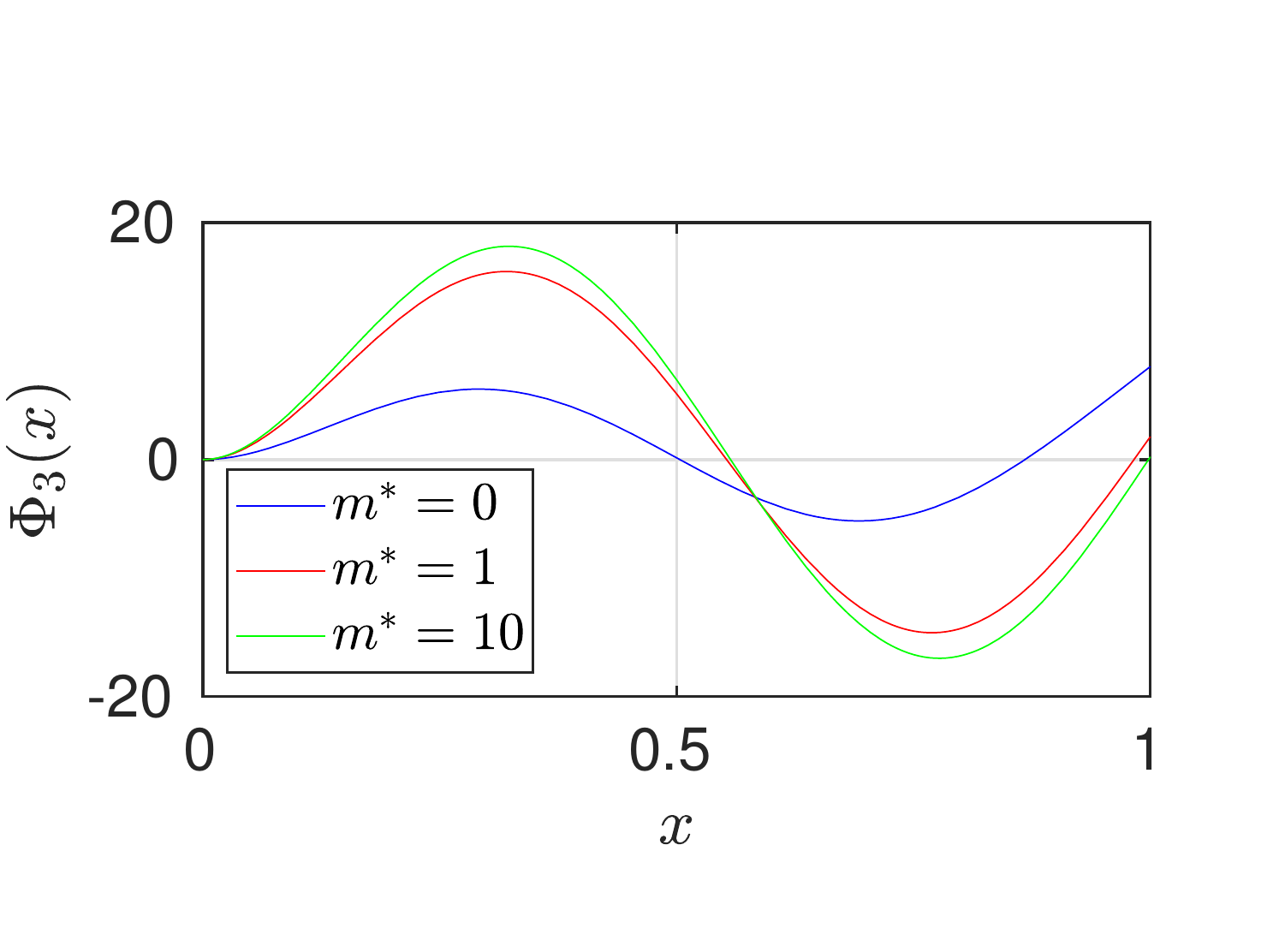}
\vspace{-0.7cm}
\caption{Influence of $m^*$ on the first three eigenmodes: (a) $\Phi_1(x)$; (b) $\Phi_2(x)$; (c) $\Phi_3(x)$. Functions normalized such that $\int_0^1 \Phi_n(x)=1$.}
\label{fig:eigenmodes}
\end{center}
\end{figure}

The equivalent spring constant associated with the free-end CNT deflection for the \textit{n-th} eigenmode, $k_n$, can be calculated as follows \cite{SMramanAPL2007}:
\begin{equation}
k_n=\dfrac{YI}{l^3}\dfrac{\int_0^1(\Phi_n''(x))^2dx}{{\Phi_n(1)}^2}.
\label{kn}
\end{equation}
By letting $I= \pi \left(d^3 g+d g^3\right)/8$ in Eq.~\ref{kn}, with $g$ and $d$ the thickness and the diameter of the CNT, respectively, the explicit form of $k_n$ becomes
\begin{multline}
k_n=\dfrac{\pi  \Omega_n^3 Y \left(d^3 g+d g^3\right)}{64 l^3 }\dfrac{ -\Omega_n \cos (2 \Omega_n)+\Omega_n\cosh (2 \Omega_n)+4 \Omega_n \sin \Omega_n \sinh \Omega_n-2 \cos \Omega_n \sinh \Omega_n+\sin (2 \Omega_n) \cosh ^2\Omega_n}{(\sin \Omega_n \cosh \Omega_n-\cos \Omega_n \sinh \Omega_n)^2}\\
+\dfrac{2 \cosh \Omega_n \left(\sin \Omega_n-\cos ^2\Omega_n \sinh \Omega_n\right)}{  (\sin \Omega_n \cosh \Omega_n-\cos \Omega_n \sinh \Omega_n)^2}.
\label{knfull}
\end{multline}

The effect of the added particle at the free end of the CNT on the standard deviation equation can be now obtained using the equipartition theorem: 
\begin{equation}
{\sigma_n}^2=\dfrac{k_B T}{k_n}.
\label{sigman}
\end{equation}
The expression in the special case of $m^*=0$ reduces to Eq.27 of \cite{SMKrishnan1998}:
\begin{equation}
{\sigma_n}^2=\frac{32 k l^3 T }{\pi  Y \left(d^3 g+d g^3\right) {\Omega_n}^4 }.
\label{reduced}
\end{equation} 
The resultant standard deviation $\sigma$ of the cantilever can be obtained by summing up all the  independent  contributions of eigenmodes. Considering the first 10 flexural modes the expression becomes: 
\begin{equation}
{\sigma}^2=\sum_{n=1}^{N=10}{\sigma_n}^2=0.84879167978\dfrac{ k_B l^3 T}{\left(d^3 g+d g^3\right) Y}.
\label{sum}
\end{equation}
The numerical coefficient in Eq.~\ref{sum} is function of the number of modes considered in the summation  $N$ and depends on the influence of the added mass. This is illustrated in Figure~\ref{fig:sigma} and reported in Tab.~\ref{tab:sigma}. The standard deviation of the cantilever is primarily given by that of the fundamental eigenmode independently of the particle mass at the free end. 

\begin{figure}[h]
\begin{center}
\includegraphics[width=0.35\linewidth]{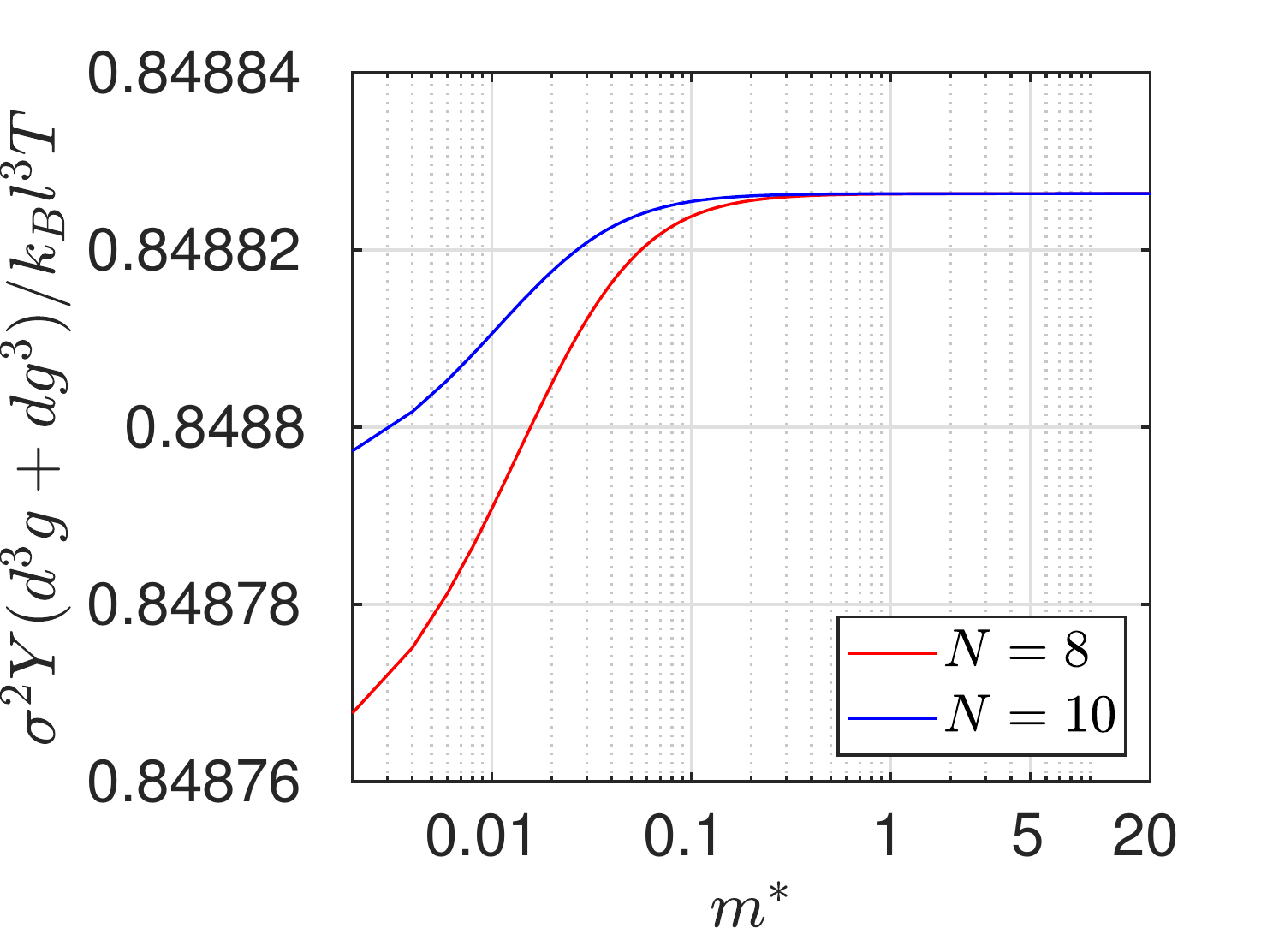}
\vspace{-.3cm}
\caption{Variation of the coefficient of  Eq.~\ref{sum} as a function of $m^*$ and while considering a different number of modes $N$ in the summation.}
\label{fig:sigma}
\end{center}
\end{figure}

\begin{table}[hb]
\caption{\label{tab:sigma}   Numerical values for the coefficient of the standard deviation in Eq.~\ref{sum} when taking into account the fundamental eigenmode only ($N=1$, middle column) and the first ten eigenmodes ($N=10$, right column).}
\begin{ruledtabular}
\begin{tabular}{ccc}
$m^*$ &${\sigma_1}^2\dfrac{Y\left(d^3 g+d g^3\right)}{k_B l^3 T}$ &$\displaystyle \sum_{n=1}^N{\sigma_n}^2\dfrac{Y\left(d^3 g+d g^3\right)}{k_B l^3 T}$\\
\hline
0&     0.8239457176           &       0.8487916797\\
0.1&  0.8363888890   &  0.8488254732    \\
0.2&  0.8414352614   &   0.8488261068   \\
0.5&  0.8462500190   &  0.8488263184  \\
1&     0.8479201355   & 0.8488263516  \\
2&     0.8485513611 &    0.8488263602 \\
5&     0.8487764942   &     0.8488263627  \\
10&   0.8488133432  &    0.8488263630   \\
20&   0.8488230357  &    0.8488263631   \\
\end{tabular}
\end{ruledtabular}

\end{table}

\subsection{Experiment}
The model of the previous subsection indicates that the platinum particle does not affect the restoring force nor the eigenmode shape of the two fundamental eigenmodes, which are polarized in perpendicular directions, while the shapes of the higher frequency eigenmodes are strongly modified by the platinum particle. For the higher frequency eigenmodes, the displacement amplitude at the free end is suppressed to zero when the particle has a larger mass than the nanotube. For this reason, our detection method based on the reflection at the free end can only measure the two fundamental modes. This is what we observe in Fig.~\ref{fig:wide_supl} for device~A. The resonances of the fundamental mode doublet are clearly visible, whereas the resonance frequencies of the second bending mode doublet are expected to be about \SI{900}{\kilo\hertz} but cannot be detected.

\begin{figure}[th]
\begin{center}
\includegraphics[width=0.9\linewidth]{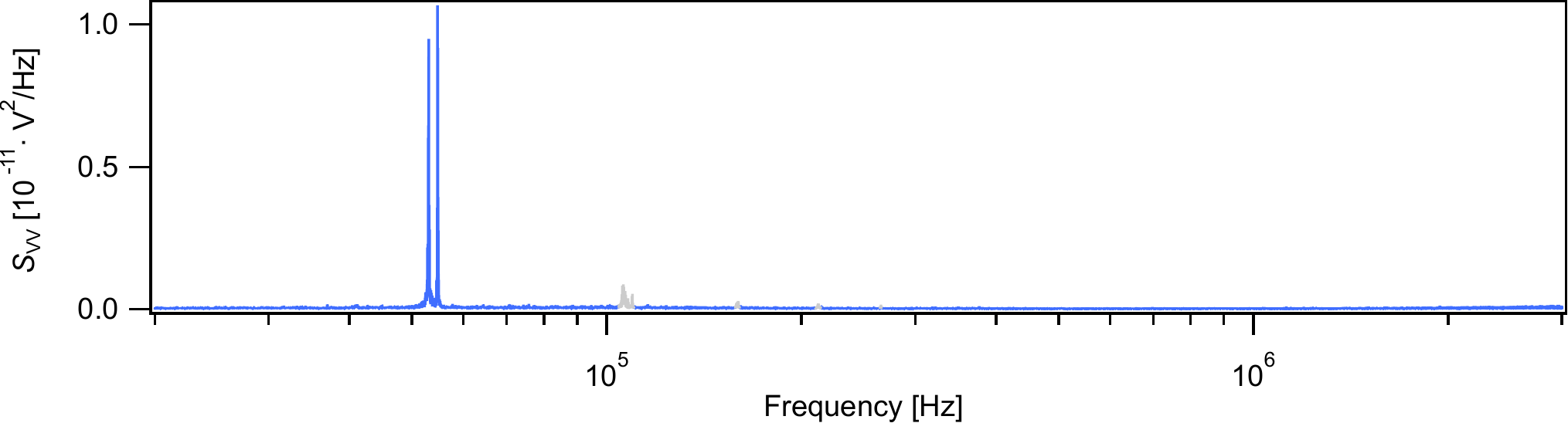}
\vspace{-.2cm}
\caption{Power spectrum of the optical reflection from device A undergoing thermal motion at \SI{300}{K}. The two near-degenerate peaks are associated with the fundamental modes polarized in perpendicular directions. The spectrum is shown over a narrower frequency range in Fig.~1c of the main text. The nonlinearity in the detection results in higher harmonics of these modes, which are marked in gray.}
\label{fig:wide_supl}
\end{center}
\end{figure}

\section{Estimation of $Y$}
\label{sec:absY}

We determine the Young’s modulus at $T=\SI{300}{\kelvin}$ for various nanotube cantilevers. We use the geometrical parameters determined by SEM and HRTEM as described in Sec.~\ref{sec:Fabrication} (see table \ref{tab:TEM}). 
The spring stiffness $k$ of a nanotube cantilever composed of $N$ concentric shells is the sum of the spring constant $k_i$ of each shell,
\begin{equation}
\label{eq:ksum}
k=\sum_{i=1}^N k_i,
\end{equation}
where we assume that the interaction between the  concentric shells has negligible contribution to the spring stiffness. We determine $Y$ of a nanotube cantilever with $N$ shells from its measured spring stiffness, its length, and the diameter $d_i$ of each shell, using
\begin{equation}
\label{eq:Yeff}
Y=0.8488\frac{kl^3}{\sum_{i=1}^N(d_i^3g+g^3d_i)},
\end{equation}
where we assume that all the shells have the same Young’s modulus and the wall thickness is $g=\SI{0.34}{\nano\meter}$. This expression can be obtained from Sec~\ref{sec:eigenmode}.

Figure \ref{fig:YTEM} shows the resulting $Y$ for the six measured devices plotted as a function of the cantilever length. The error bars represent the standard error $\Delta Y$ for each measurement, which is determined by expanding Eq.~\ref{eq:Yeff} and calculating the propagation of the measurement uncertainties in $l$, $k$ and $d$ (see table \ref{tab:TEM}). The solid line is the mean Young's modulus $\bar{Y}=\sum Y/N=\SI{1.06}{\tera\pascal}$ whereas the dashed lines indicate the confidence intervals $\Delta \bar{Y}$. The latter is estimated by summing the standard error of the $Y$ value of the different cantilevers and the mean of their standard error $\Delta Y$ divided by $\sqrt{N}$, which yields $\Delta \bar{Y}=\pm\SI{0.28}{\tera\pascal}$.

\begin{figure}[th]
\begin{center}
\includegraphics[width=0.5\linewidth]{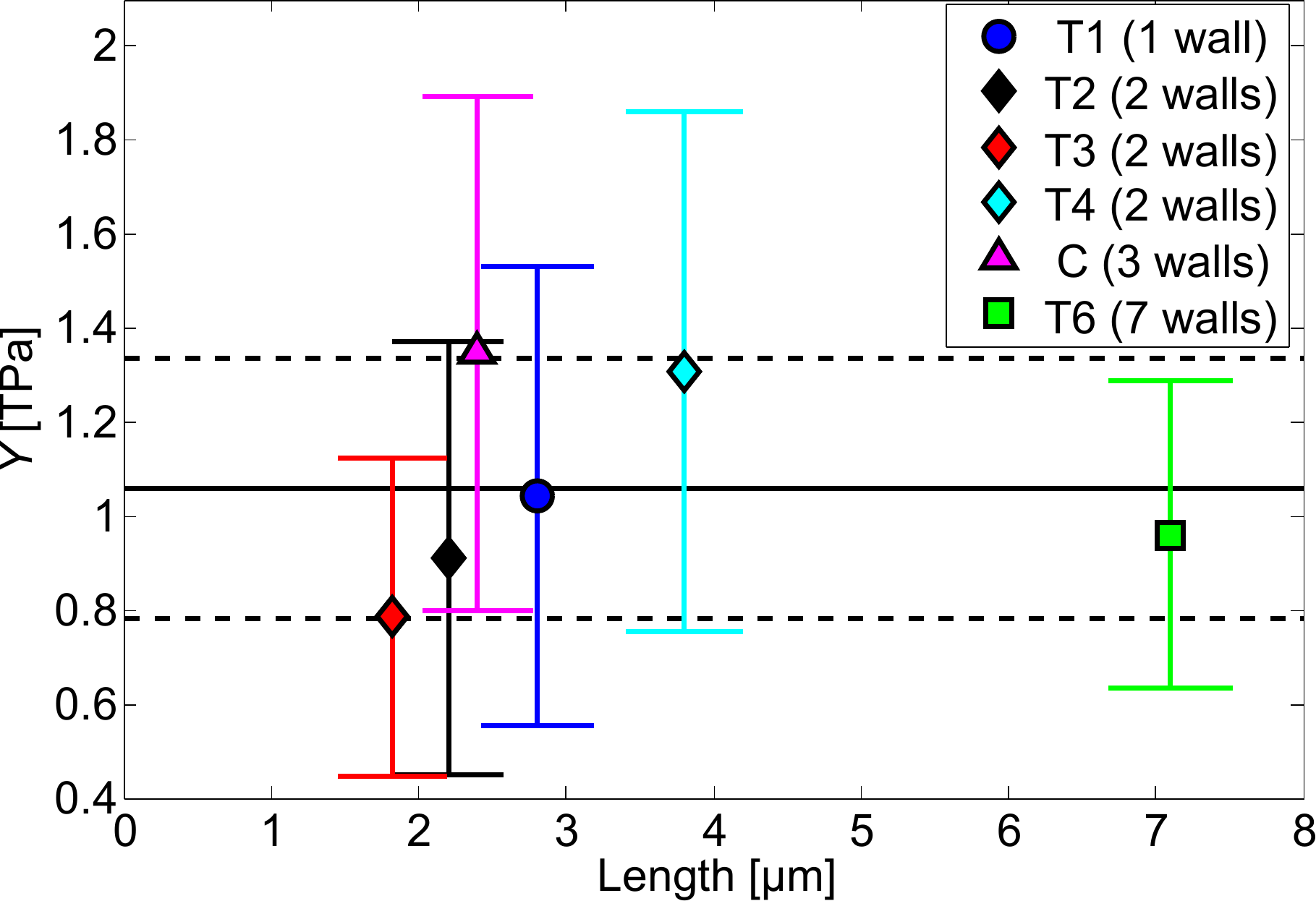}
\vspace{-.2cm}
\caption{$Y$ determined from SEM and HRTEM for six different nanotube cantilevers. The error bars represent the standard error for each measurement. The solid black line marks the mean value $\bar{Y}=\SI{1.06}{\tera\pascal}$ of all the measurements and the dashed black lines indicate the corresponding confidence intervals $\Delta \bar{Y}=\pm\SI{0.28}{\tera\pascal}$.}
\label{fig:YTEM}
\end{center}
\end{figure}

\section{Discussion on the origin of the temperature dependence of the resonance frequency}

In the main text, we discuss the measured temperature dependence of the resonance frequency in terms of the variation of $Y$. Here, we consider other possible mechanisms but we show that they cannot account for our measurements.

The measured $T$ dependence of $\omega_0$ could originate from the variation of the mass $m$ adsorbed on the nanotube. However, mass adsorption, which occurs when lowering $T$, would lead to a reduction of $\omega_0$ \cite{SMGruber2019,SMWang2010,SMYang2011,SMTavernarakis2014}, which is just the opposite of what is measured. Moreover, we do not observe any hysteresis in $\omega_0$ when cooling the device from \SI{300}{K} to cryogenic temperatures and then heating it back to \SI{300}{K}, which shows that temperature-induced mass adsorption and desorption plays a negligible role \cite{SMTavernarakis2014}. Thus, the measured variation of $\omega_0(T)$ is not accounted for by adsorbed mass changes. 

\begin{figure}[b]
\centering
\includegraphics[width=0.5\linewidth]{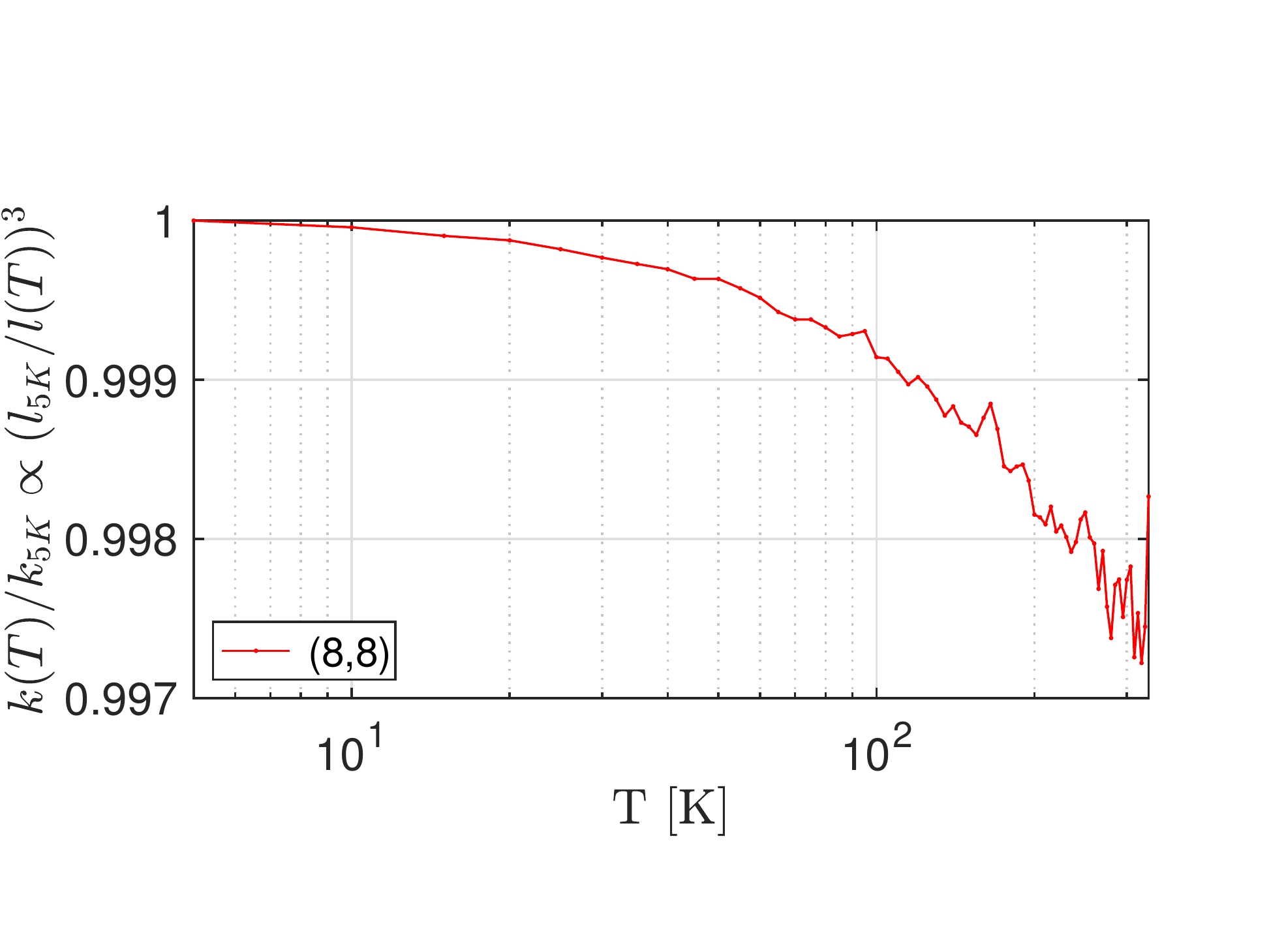}
\vspace{-1cm}
\caption{Predicted relative change in stiffness with respect to temperature induced by the nanotube elongation. MD simulations for a (8,8) CNT fully clamped at one end and free on the other.}
\label{fig:stffnsselongation}
\end{figure} 

The measurements could be related to the change in the length of the nanotube when the thermal environment is varied, since the spring constant depends on the nanotube length as $k \propto l^{-3}$. However, the measured resonance frequency reduction at room temperature $\Delta \omega_0(T=300~K)/\omega_0\simeq -2.2\times10^{-2}$ for device~A is much larger than the predicted reduction $\Delta \omega_0(T=300~K)/\omega_0=-1.25\times10^{-3}$ based on the longitudinal expansion of the nanotube in different thermal conditions obtained from our molecular dynamics simulations for a  (8,8) CNT. 
The predicted relative change in stiffness $k(T)/k_{5K}$ associated with the thermal expansion of the nanotube as a function of temperature is shown in Fig.~\ref{fig:stffnsselongation}. The results are obtained by calculating the elongation of the nanutube for different thermalisation temperatures.
Here we assume that the stiffness ratio $k(T)/k_{5K}$ is proportional to the cube of the function $(l_{5K}/l(T))$ in which $l_{5K}$ and $l(T)$  are the length of the CNT at $5K$ and at  temperature $T$, respectively. The decreasing behaviour reported in Fig.~\ref{fig:stffnsselongation} suggests that the nanotube stretches with the increase in temperature. Overall, this shows that the thermal expansion is not the cause of the measured $\omega_0(T)$ reduction.

Another possible origin could be the nanotube resonance frequency change that arises from the combination of the Duffing nonlinearity and the thermal motion. Figure~\ref{fig:duffing}a shows the measured variation of the resonance frequency as a function of driven vibrational amplitude $\left<x_{\textrm{vibra}}^2\right>$, which allows us to quantify the Duffing constant $\gamma_\textrm{eff}$ using
\begin{equation}
\Delta\omega=\dfrac{3}{8}\dfrac{\gamma_\textrm{eff}}{\omega_0}\left<x_{\textrm{vibra}}^2\right>.
\label{eq:duffing}
\end{equation}
The driven amplitude is calibrated following the procedure described in Ref.~\cite{SMTavernarakis2018}. We compute the linear temperature dependence of the resonance frequency expected from the combination of the Duffing nonlinearity and the thermal vibrations using
\begin{equation}
\Delta\omega=\dfrac{3}{8}\dfrac{\gamma_\textrm{eff}}{\omega_0}\left<x_{\textrm{th}}^2\right>=\dfrac{3}{8}\dfrac{\gamma_\textrm{eff}}{\omega_0}\frac{k_\textrm{B}T}{k}.
\label{eq:duffingthermal}
\end{equation}
Figure~\ref{fig:duffing}b shows that the slope of the expected dependence is positive, in contrast to what we measure. Moreover, the frequency shift $\Delta \omega_0(T=300~\textrm{K})/\omega_0=3.0\times10^{-5}$ is much smaller in magnitude than the measured value $\Delta \omega_0(T=300~\textrm{K})/\omega_0\simeq -2.2\times10^{-2}$. This shows that the Duffing nonlinearity together with the thermal vibrations cannot describe our experimental findings.

Another explanation for our data could be related to the diffusion of adsorbed atoms along the nanotube. Mechanical  vibrations lead to a force that pushes atoms towards the anti-node of the mode \cite{SMAtalaya2011}. Enhancing the vibrational amplitude of the fundamental mode results in more atoms near the nanotube free end and, therefore, a larger effective mass of the mode and a lower resonance frequency. However, we observe the opposite behaviour in Fig.~\ref{fig:duffing}a. This shows that the effect of the diffusion of adsorbed atoms is smaller than that of the Duffing nonlinearity, so that it cannot account for the measured $T$ dependence of $\omega_0$.

\begin{figure}[hb]
\centering
\includegraphics[width=0.95\linewidth]{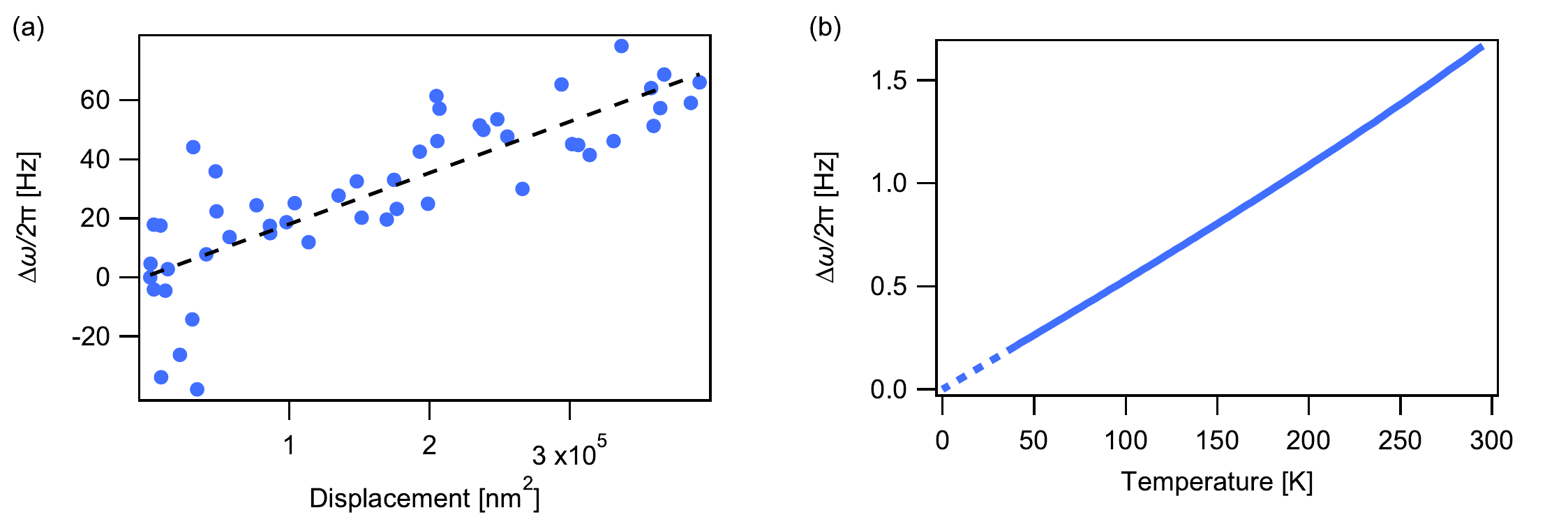}
\vspace{-.2cm}
\caption{Estimation of the Duffing constant and its effect on the temperature dependence of the resonance frequency for device~A. (a) Variation of the resonance frequency as a function of driven vibrational amplitude measured at $T=\SI{100}{\kelvin}$. (b) Estimated frequency shift as a function of temperature due to the Duffing constant and the thermal vibrations.}
\label{fig:duffing}
\end{figure}

\section{Molecular dynamics simulations}

We report molecular dynamics simulations of the  Brownian motion of carbon nanotubes over a finite temperature range.
Simulations are carried out in the Large-scale Atomic/Molecular Massively Parallel Simulator (LAMMPS) software \cite{SMplimpton2007lammps} for single layer CNTs of different chirality. In Figure ~\ref{fig:MDrender} we showcase the geometry of one such CNT.

To account for atom-atom interactions, we use the Tersoff potential \cite{SMTersoffPRB1988} with optimized parameters for lattice dynamics and phonon thermal transport
\cite{SMLindsayPRB2010}. We note that this potential is commonly used for simulating atomic interactions and predicting mechanical properties of  carbon-based nanomaterials \cite{SMTangPRB2009,SMYakobsonPRL1996}.

\begin{figure}[t]
\centering
\includegraphics[width=0.7\linewidth]{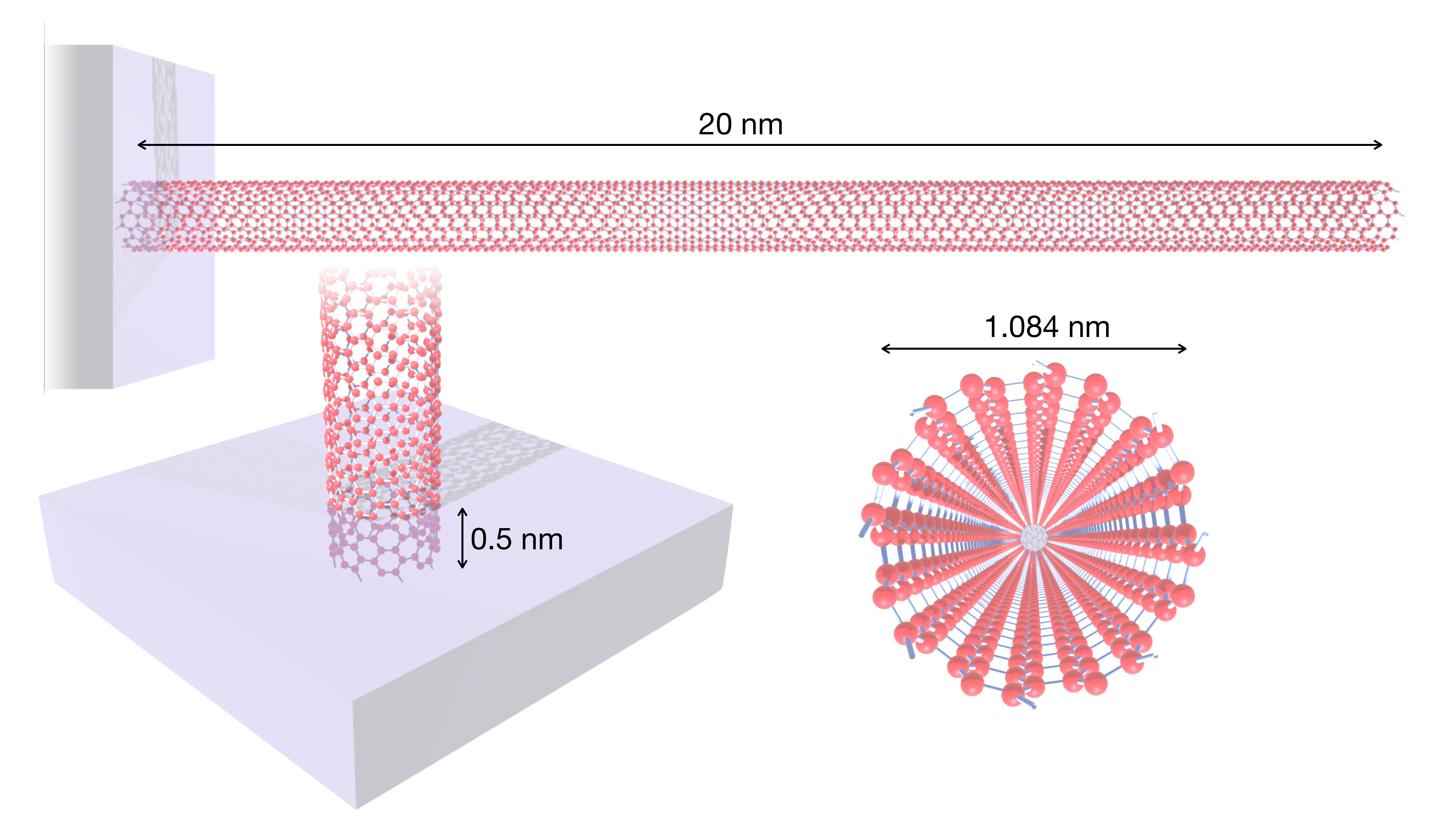}
\caption{(8,8) CNT with a total of 2624 atoms. Atoms at one end are clamped for a length of 0.5 nm. The (8,8) CNT has a radius of 0.542 nm and a total length of 20 nm.}
\label{fig:MDrender}
\end{figure}

 To track the Brownian motion of the nanotube, the system is initially relaxed to ensure  equilibrium at the minimum potential state.
 The minimization of the total potential energy is performed via the Polak-Ribiere conjugate gradient algorithm \cite{SMKlessig}. 
 The starting point for the minimization procedure is the initial configuration of the atoms, and the potential energy of the system is considered to be in a local  minimum when its energy is less than $1\times 10^{-10}$ eV or when the forces are less than $1\times 10^{-10}$ eV/\AA.
After the relaxation, at one end, the translational degrees of freedom are constrained for all atoms for a length of 5 nm (see Figure~\ref{fig:MDrender}). This constraint is applied to obtain a CNT cantilever. Once the equilibrium position is obtained, Newton's equations are  integrated using the velocity-Verlet  algorithm, with a time-step $dt=0.1$ fs to determine the variation of the position and velocity  of the atoms.

To account for the thermal effects, the system is then equilibrated in a constant volume and temperature ensemble (NVT). The temperature is first brought to a certain value and then kept constant by applying the Nose-Hoover algorithm that thermostats the translational velocity of atoms \cite{SMdoi:10.1063/1.449071}.
The algorithm for the thermalisation is applied for 10 ns to ensure that a stable temperature is obtained (see Fig.\ref{fig:termalisation}).
Once thermal equilibrium is reached, the vibration response is studied in an energy conserving ensemble (NVE).
In this context, the thermal fluctuations of the CNT are monitored  for 50ns  discarding an initial transient response of 10 ns, and the coordinates of all atoms are saved every 2.5 ps. 

\begin{figure}[t]
\centering
\includegraphics[width=0.35\linewidth]{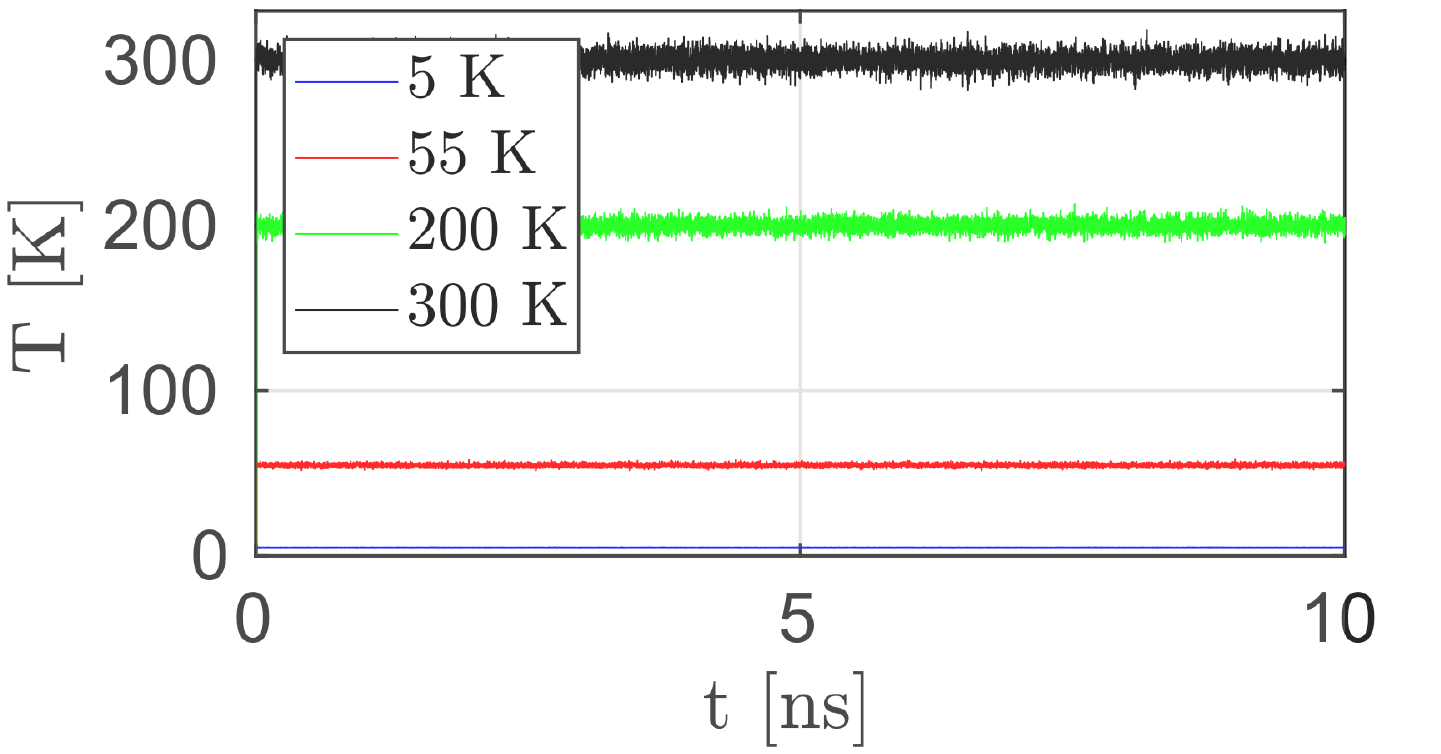}
\vspace{-.3cm}
\caption{Temperature fluctuation during the termalization phase for a (8,8) CNT. For the thermostat temperature of 300 K we obtain a mean of 300.031 K and standard deviation of 5.69 K.}
\label{fig:termalisation}
\end{figure}

To obtain the resonance frequencies of the CNT, we compute the FFT of the extracted time signals from molecular dynamics. An example of one such FFT averaged over all atoms is shown in Figure~\ref{fig:MDfft}(a) for a (8,8) CNT at 50 K. The thermal influence on the mechanics of the CNT is obtained by tracking the natural frequencies as a function of the thermostat temperature. The relative change of the square of the frequencies for the first three flexural modes for the (8,8) CNT cantilever is shown in Figure~\ref{fig:MDfft}(b); this quantity is equal to the relative change of the spring constant $\Delta k(T)/k$ .
The first three flexural modes highlight the same reduction with respect to the variation of the thermal bath. The staircase behaviour of the first flexural mode is due to the insufficient resolution in frequency (i.e. 20 MHz). Our results are compared to experimental measurements in Fig.~3 of the main text.

\clearpage
\begin{figure}[ht]
\centering
\includegraphics[width=0.45\linewidth]{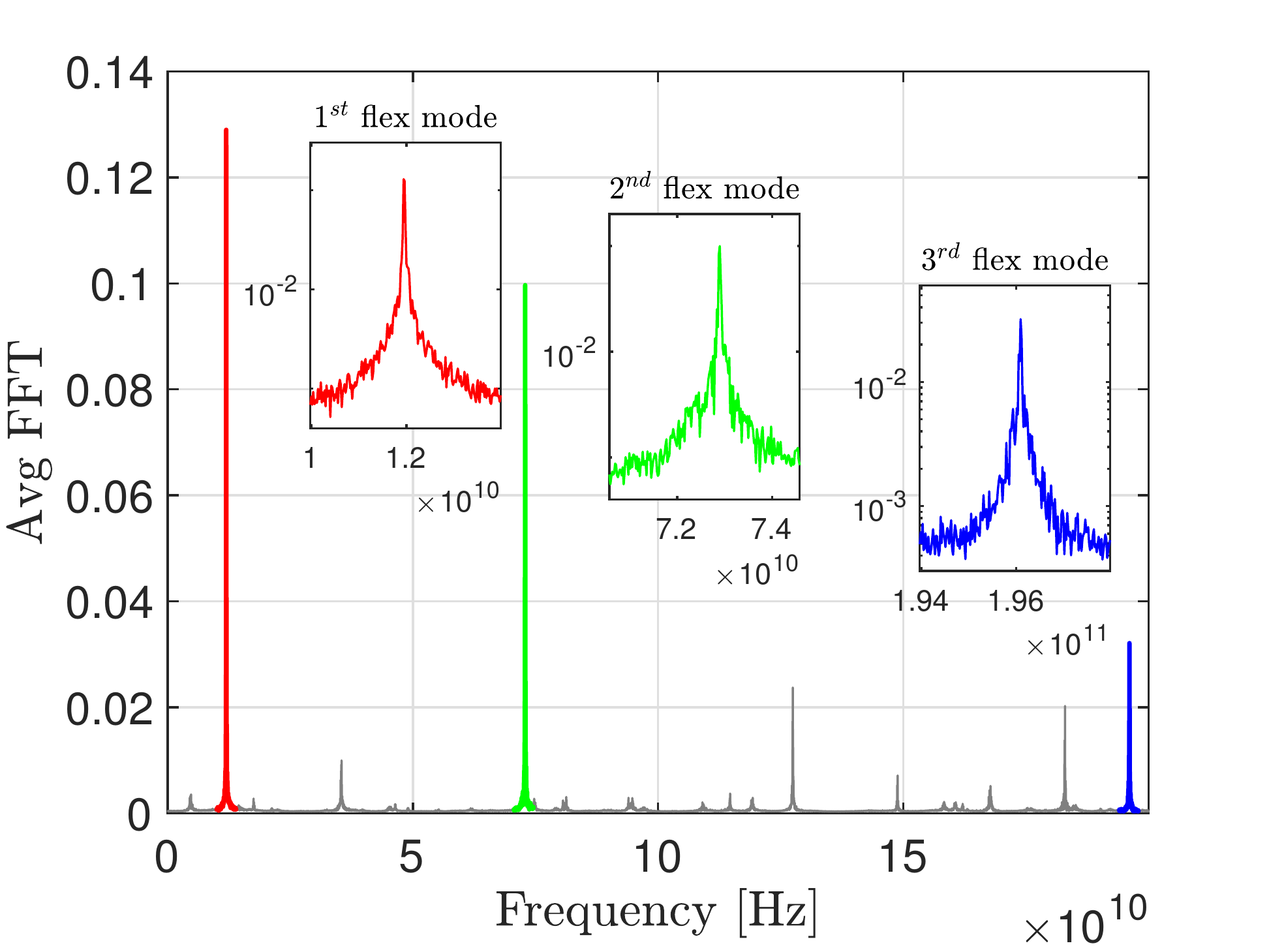}
\includegraphics[width=0.45\linewidth]{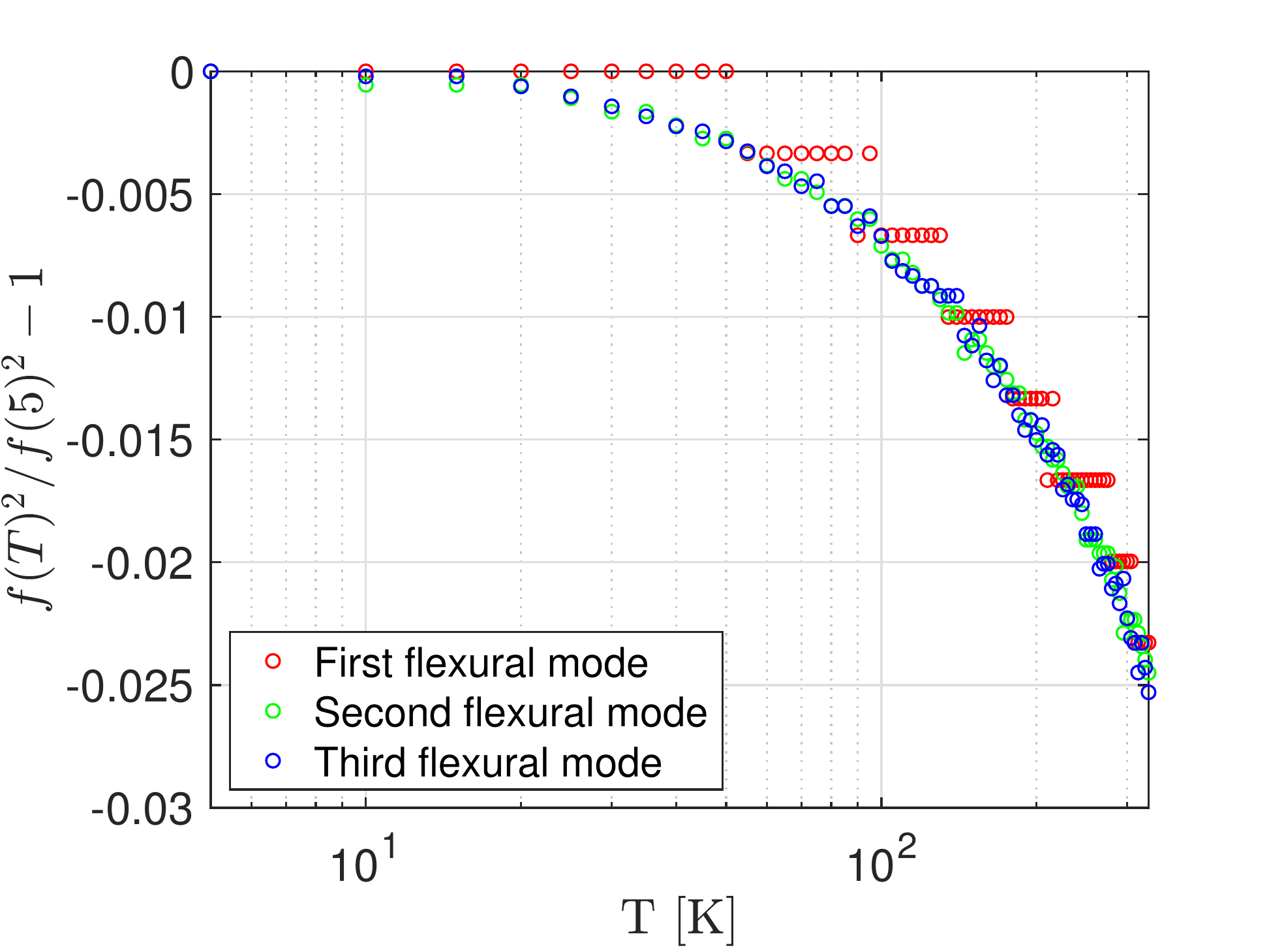}
\vspace{-.4cm}
\caption{a) The averaged frequency spectrum of all atoms for the (8,8) CNT at $T=50$ K. . b) Relative change of the square of frequency with temperature for the first three flexural modes of the CNT from 5K to 330 K with a temperature increment of 5 K. The staircase behaviour of the first flexural mode is due to the insufficient resolution in frequency}
\label{fig:MDfft}
\end{figure}

We remark that the spectral analysis performed to extract the thermal behaviour of the system, does not allow  for an immediate classification of the natural modes of the system and their associated eigenfrequencies.
However, it is possible to unravel spatial information of the nanotube from the time response data via the proper orthogonal decomposition (POD) method. The details of this technique can be found in \cite{SMsajadi2019nonlinear} and are briefly described in Sec.~\ref{SI:POD}. Using POD, we can identify the eigenmodes corresponding to the resonance peaks of Figure~\ref{fig:MDfft}(a). 
The  mode shapes for the first three flexural modes  obtained via POD for a (8,8) CNT at 50 K are reported in Figure~\ref{fig:MDmodes}.
The procedure outlined above has been repeated in the temperature range $T\in[5,330]$ K, for three different chiralities namely (5,10), (8,8),and (10,10), and the results are shown in Figure 3(b).
 
\begin{figure}[h]
\centering
\includegraphics[width=0.30\linewidth]{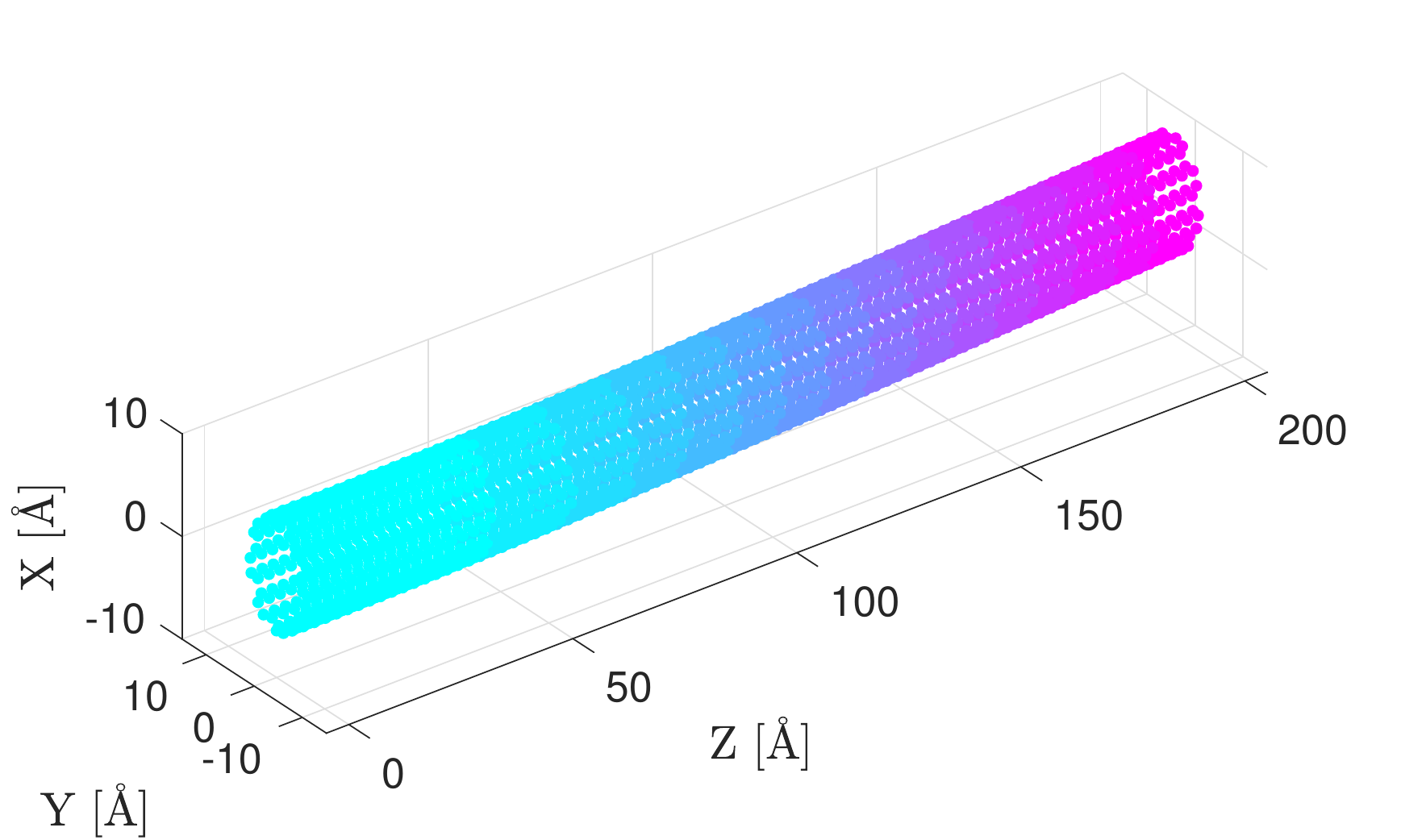}
\includegraphics[width=0.30\linewidth]{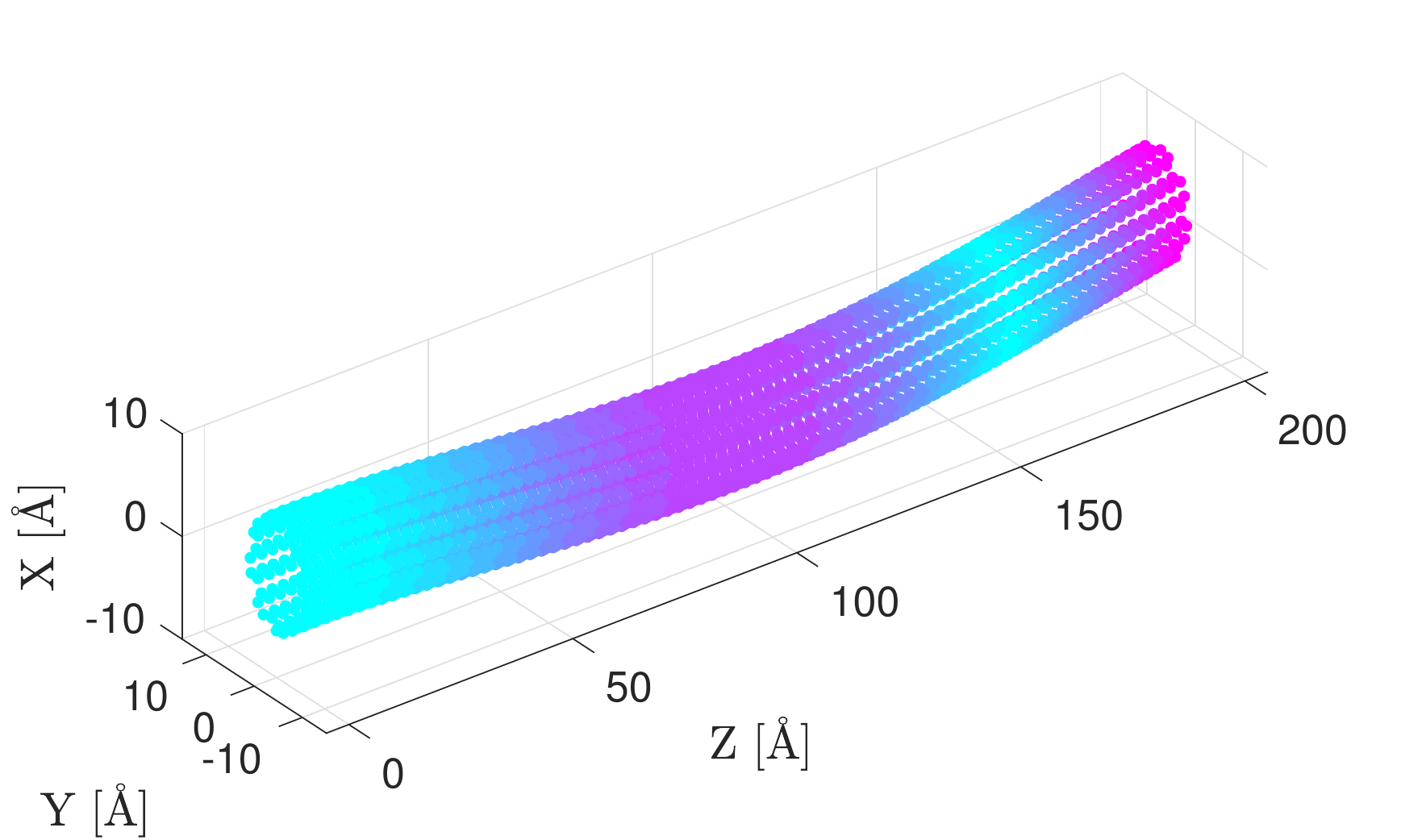}
\includegraphics[width=0.30\linewidth]{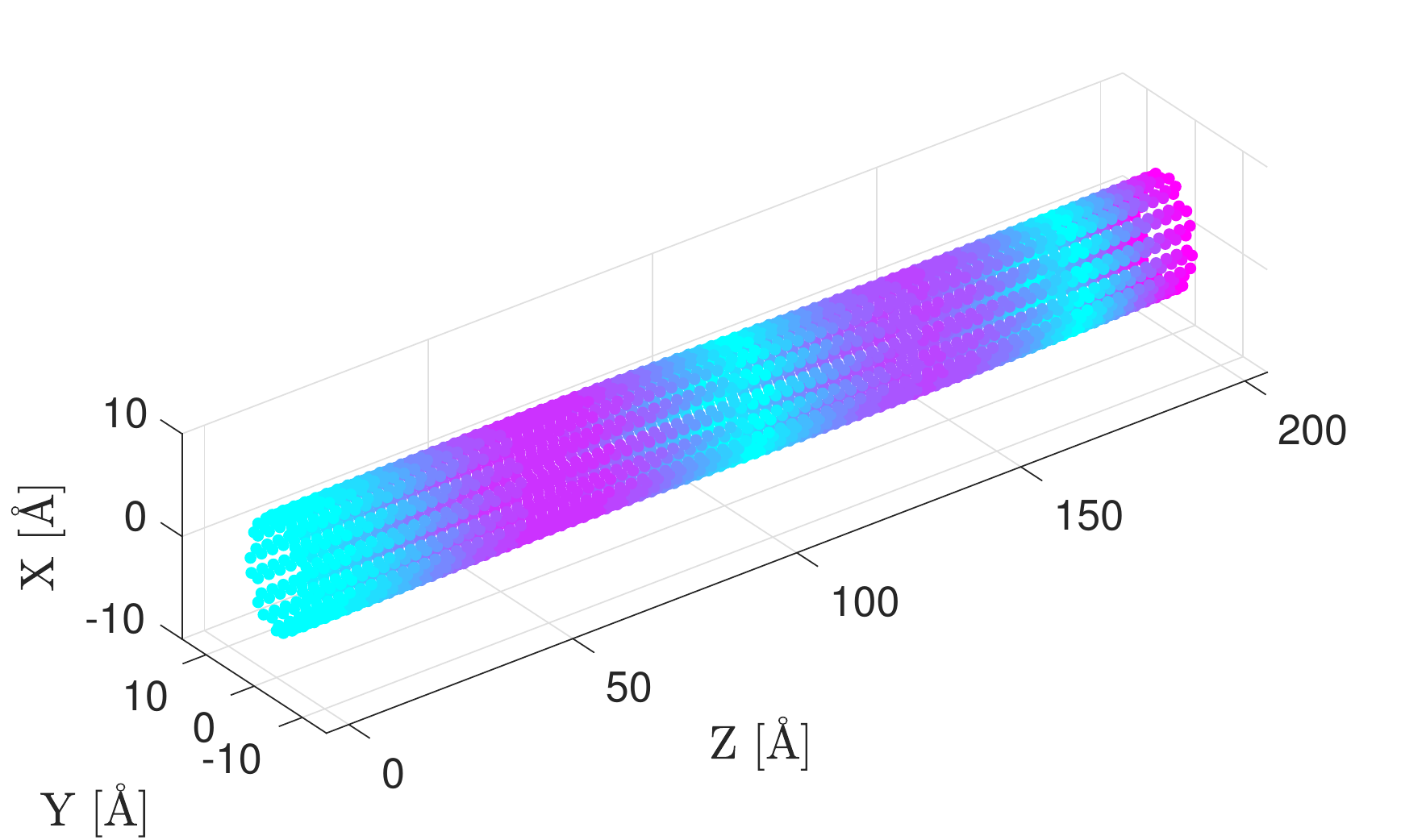}
\vspace{-.2cm}
\caption{Mode shapes obtained via proper orthogonal decomposition at $T=50$ K. (a) First flexural mode. (b) Second flexural mode. (c) Third flexural mode. 
Modes are amplified ten times for visualization. 
Colormap for the norm of x and y displacement.}
\label{fig:MDmodes}
\end{figure} 
 
\subsection{Proper orthogonal decomposition}
\label{SI:POD}

The MD simulations provide the time response in a vector $\mathbf{u}$ comprising the position of $M-$atoms. 
The time history consists of N snapshots of the motion as $\pqpq{\mathbf{u}(t_1),\mathbf{u}(t_2), \ldots,\mathbf{u}(t_N)}$. We remove the time average (mean values) of the responses by obtaining the time-varying part, $\mathbf{x}(t_i) = \mathbf{u}(t_i)-\text{mean}(\mathbf{u})$.
To extract the proper orthogonal modes of vibrations, a discrete matrix $\mathbf{X}$ is first built such that each row corresponds to a time response of one atom and each column corresponds to a snapshot of the CNT at a specific time as:

\begin{equation}
\mathbf{X}=\left[\begin{matrix}\mathbf{x}(t_1)& \mathbf{x}(t_2)& \cdots  &   \mathbf{x}(t_N)\end{matrix} \right]=\left[\begin{matrix}
 x_1(t_1)  &  \cdots  &  x_1(t_N)  \\ 
\vdots  &  \ddots & \vdots  \\ 
 x_M(t_1)  &  \cdots &  x_M(t_N) 
\end{matrix} 
\right] ,
\label{matrixX}
\end{equation}
where $x_i(t_j)$ is the response of the $i-th$ atom at time $t_j$.
Once matrix $\mathbf{X}$ is constructed, the orthogonal modes are obtained by using the singular-value decomposition (SVD) of the discrete matrix. The SVD operator decomposes $\mathbf{X}$ as:
\begin{equation}
\mathbf{X}=\mathbf{U}\Sigma\mathbf{V^*},
\label{SVD}
\end{equation}
where $\mathbf{U}$ is an $M\times M$ real or complex unitary matrix,
$\Sigma$ is a $M\times N$ rectangular diagonal matrix with non-negative real diagonals $\sigma_i$ that are the singular values of $\mathbf{X}$,
and $\mathbf{V}$ is an $N\times N$ real or complex unitary matrix, with $\mathbf{V^*}$ being its conjugate transpose. 
The columns of $\mathbf{U}$ and $\mathbf{V}$ are the so-called  left-singular and right-singular vectors of $\mathbf{X}$, respectively. 
Among these matrices,  $\mathbf{U}$ corresponds to proper orthogonal modes of vibration that can linearly obtain all the snapshots of the motion with minimum error. Using this matrix we can identify  the modes corresponding to the peaks seen in Figure~\ref{fig:MDfft} and report them in Figure~\ref{fig:MDmodes}.

\section{Quasi-harmonic approximation}

The elastic constants of a solid are defined as appropriate derivatives of the free energy with respect to strain tensor components~\cite{SMLandau}. In particular, the Young’s modulus of a nanotube along the axial direction can be computed from:
\begin{equation}
Y(T) = \frac{1}{V_0(T)} \left( \frac{\partial^2 F(T,\epsilon)}{\partial \epsilon^2} \right)_{\epsilon=0},
\label{eq:YoungT}
\end{equation}
where $F(T,\epsilon)$ is the free energy at temperature~$T$ and strain~$\epsilon$, and $V_0(T)$ is the equilibrium volume at that temperature. It is frequent to assume that the temperature dependence of elastic constants is small and close to their zero-temperature value, which amounts to substituting the internal energy in place of the free energy in Eq.~(\ref{eq:YoungT}). However, in this work we are particularly interested in the temperature-dependence of the Young's modulus. To this end we resort to a quasi-harmonic approximation of the free energy:
\begin{equation}
F(T,\epsilon) \approx F_0(\epsilon) + F_{vib}(T,\epsilon) = F_0(\epsilon) +  k_\textrm{B} T \sum_{n{\bf k}} \ln \left[ 2 \sinh \left(\frac{\hbar \omega_{n{\bf k}}(\epsilon)}{2 k_\textrm{B} T}\right)\right],
\label{eq:HelmholtzFree}
\end{equation}
where $F_0(\epsilon)$ is the free energy at zero temperature (i.e. the potential energy) at $\epsilon$~strain, $\omega_{n{\bf k}}(\epsilon)$ is the frequency of vibrational mode~$n$ at reciprocal lattice vector~${\bf k}$ calculated at strain $\epsilon$. The nanotube phonon frequencies have been calculated using a tight-binding model~\cite{SMdftb} employing the PHON package~\cite{SMphon} (Fig.~\ref{fig:eduardo}). From the free energy we can obtain the temperature-dependent Young's modulus via Eq.~(\ref{eq:YoungT}). Our results are compared with experimental measurements in Fig.~3 of the main text. 

\begin{figure}[h]
\centering
\includegraphics[width=0.48\linewidth]{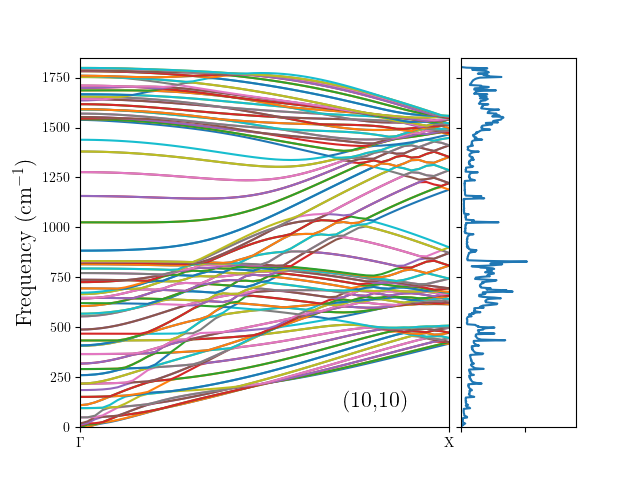}
\includegraphics[width=0.48\linewidth]{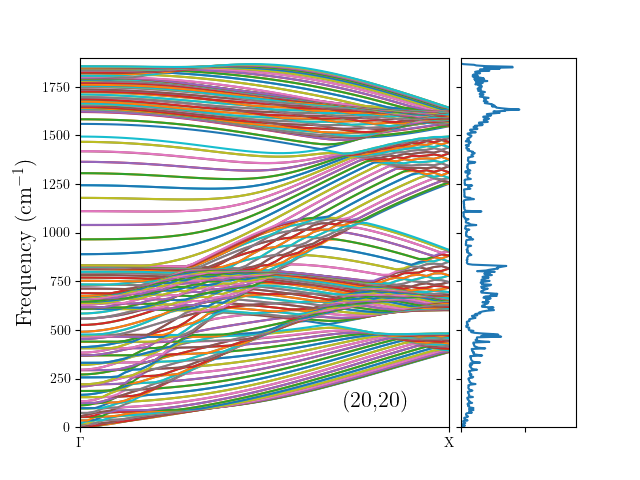}
\vspace{-.5cm}
\caption{(a) panel shows the phonon band structure and vibrational density of states for the (10,10) nanotube (see text). (b) panel shows the same information for the (26,0) nanotube.}
\label{fig:eduardo}
\end{figure}

\section{Estimation of phonon decay rates}
\label{sec:lifetime}

We estimate the decay rates for different phonon modes using the expressions derived by de Martino et al.\cite{SMdemartino2009}. For the longitudinal phonon modes the decay rate due to phonon-phonon interactions is given by 

\begin{equation}
\tau_\textrm{L}^{-1} = \frac{\hbar}{4\pi \rho r^4}\left\{\frac{\sqrt{k_\textrm{ph}r}}{2^{5/4}}\textrm{coth}\left(\frac{\hbar v_\textrm{L}k_\textrm{ph}}{4k_\textrm{B}T}\right)+ \sqrt{2}\textrm{exp}\left(\frac{-\hbar v_\textrm{L}}{2\sqrt{2}rk_\textrm{B}T}\right)\textrm{sinh}\left(\frac{\hbar v_\textrm{L}k_\textrm{ph}}{2}\right)\right\}
\label{eq:tauL}
\end{equation}
where $\hbar$ is the reduced Planck constant, $\rho=\SI{3.8e-7}{\kilogram/\meter^2}$, $r$ is the nanotube radius, $k_\textrm{ph}$ is the phonon wave number, and $v_\textrm{L}=\SI{1.99e4}{\meter/\second}$ is the longnitudinal speed of sound. For the twist phonon modes the decay rate is

\begin{equation}
\tau_\textrm{T}^{-1} = \frac{\hbar}{2\rho}\left(\frac{v_\textrm{T}}{v_\textrm{L}}\right)^{7/2}\frac{2^{1/4}\left(k_\textrm{ph}r\right)^{3/2}}{8\pi r^4}\left\{\textrm{coth}\left(\frac{\hbar v_\textrm{T}k_\textrm{ph}}{4k_\textrm{B}T}\right)+2^{5/4}\left(\frac{v_\textrm{T}}{v_\textrm{L}k_\textrm{ph}r}\right)^{3/2}\textrm{exp}\left(-\frac{\hbar v_\textrm{T}^2}{2\sqrt{2}v_\textrm{L}rk_\textrm{B}T}\right)\textrm{sinh}\left(\frac{\hbar v_\textrm{T}k_\textrm{ph}}{2k_\textrm{B}T}\right)\right\}
\label{eq:tauT}
\end{equation}
where $v_\textrm{T}=\SI{1.23e4}{\meter/\second}$. We calculate the decay rates for different phonon energies $E_\textrm{ph}$. The wave number is $k=E_\textrm{ph}/\hbar v_\textrm{L}$ for longitudinal and $k=E_\textrm{ph}/\hbar v_\textrm{T}$ for twist phonons and $r=\SI{1}{\nano\meter}$. Figs. \ref{fig:tau} (a)-(b) show the respective temperature dependencies of $\tau_\textrm{L}$ and $\tau_\textrm{T}$. 

\begin{figure}[th]
\begin{center}
\includegraphics[width=0.45\linewidth]{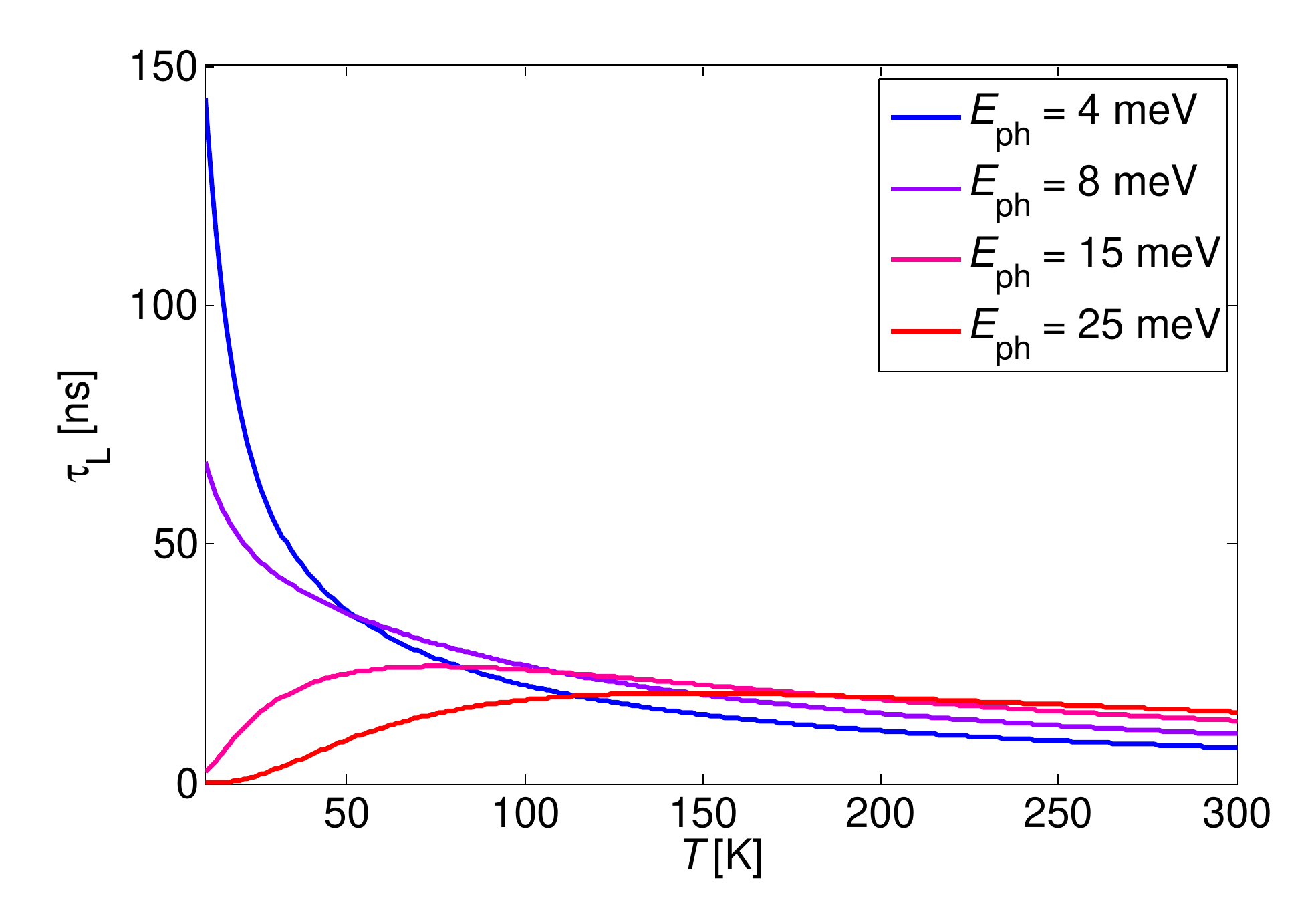}
\includegraphics[width=0.45\linewidth]{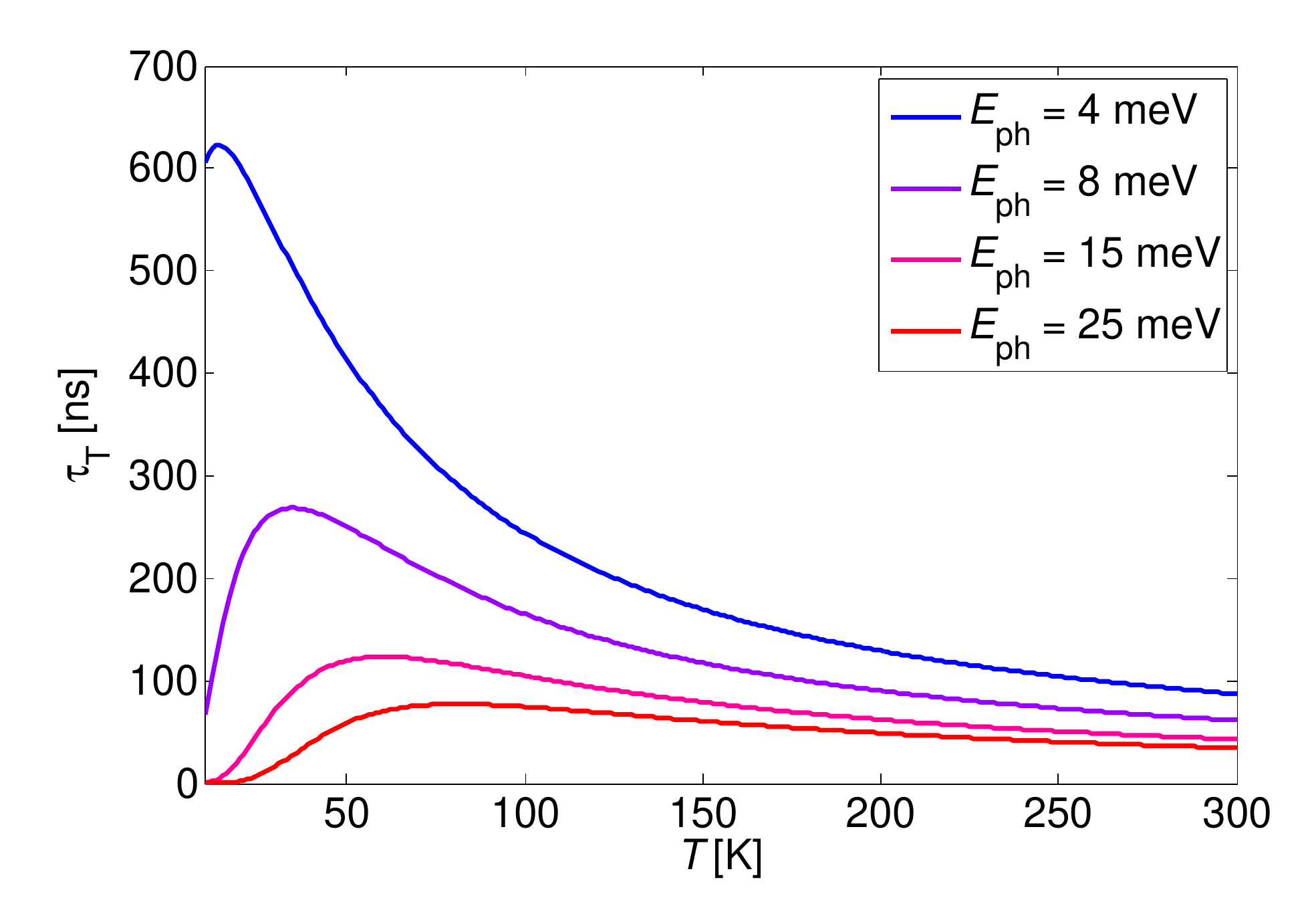}
\vspace{-.5cm}
\caption{Temperature dependence of the phonon decay times of the longitudinal (left) and twist phonon modes (right) for different phonon energies $E_\textrm{ph}$.}
\label{fig:tau}
\end{center}
\end{figure}

\section{Akhiezer dissipation }
\label{sec:akhiezer}
The general expression of the Akhiezer dissipation rate $\Gamma_{Akh}$ of a mechanical eigenmode can be found in Eq.~70 of Ref.~\cite{SMAtalaya2016}. We do not reproduce it here, since the description of the different terms would be rather long. This expression is derived for nonlinear dissipation, but the linear Akhiezer dissipation rate can be obtained from this Eq.~70 when using the proper $V_{\alpha}$ for the coupling between the mechanical resonator and the high-energy phonon modes. 

The temperature enters in $\Gamma_{Akh}$ through (\textit{i}) the number of phonon modes with energy $\hbar \omega_k \lesssim  k_\textrm{B}T$, (\textit{ii}) the thermal number $\bar n_k$ of quanta for each phonon mode, and (\textit{iii}) the decay rate $1/\tau_k$ of each phonon mode. When increasing the temperature, the contributions (\textit{i}) and (\textit{ii}) enhance $\Gamma_{Akh}$, while the contribution (\textit{iii}) lowers it. Moreover, when the decay rate of a phonon mode is much shorter than the period of the mechanical eigenmode, $\omega_0\tau_k\gg1$, this mode does not contribute to $\Gamma_{Akh}$. Since the density of states of phonons varies up and down as a function of energy (Fig.~\ref{fig:eduardo}), it might well happen that $\Gamma_{Akh}$ depends in a non-monotonic way on temperature, resulting in peaks in dissipation.

\section{Dissipation due to defects}
\label{sec:defects}

We show in this section that the measured temperature dependence of the dissipation cannot be described by the model that is used in the literature \cite{SMFaust2014,SMHamoumi2018} to quantify dissipation due to defects. Peaks in dissipation when sweeping temperature is often attributed to  microscopic defects. These defects are modelled by double-well potentials with barrier height $V_0$ and asymmetry $\Delta$ between the two wells. At the high temperature of our experiments, the passage from one well to the other well is thermally activated with a characteristic time 
\begin{equation}
\tau_d=\tau_{d0}\exp{(V_0/k_\textrm{B}T)}, 
\label{eq:taud}
\end{equation}
with $1/\tau_{d0}$ the attempt rate to overcome the barrier. Assuming that all defects have similar $V_0$, $\Delta$, and $\tau_{d0}$, a peak in dissipation occurs when the characteristic rate $1/\tau_{d}$ of the defects matches the mechanical resonance frequency, $1/\tau_d=\omega_0$. Our measurements in Fig.~4b of the main text show dissipation peaks at different temperatures. Using the values of these temperatures together with $1/\tau_d=\omega_0$, we construct a plot of $\tau_d$ as a function of $T$ (Fig.~\ref{fig:TLS}a). Despite the relatively large spread in the values of $\tau_d$ in Fig.~\ref{fig:TLS}a, the data cannot be described by an exponential behaviour, suggesting that $\exp{V_0/k_\textrm{B}T}\sim1$ in the measured temperature range in order to force a reasonable description of the data by Eq.~\ref{eq:taud}. Such an analysis would lead to an unrealistically long $\tau_{d0}\sim2\times10^{-6}$~s, considering that $\tau_{d0}$ is typically in the  $10^{-13}$~s -- $10^{-11}$~s range \cite{SMFaust2014,SMHamoumi2018,SMVacher2005}. If we were considering two or three different types of defects, each of them with well defined characteristics $V_0$, $\Delta$, and $\tau_{d0}$, we would also obtain $\tau_{d0}$ in the $\SI{}{\micro\second}$ range. Therefore, the model based on defects with narrow characteristics distribution cannot account for our measurements. Another possibility with the double-well potential model is to assume a broad distribution of the defect characteristics $V_0$ and $\Delta$ \cite{SMFaust2014,SMHamoumi2018,SMVacher2005}. A peak in dissipation can be obtained in a specific parameter space region. The peak always features a negative curvature between $T=0$~K and the peak temperature (Fig.~\ref{fig:TLS}b), which is just the opposite of what is observed in our experiments (Fig.~4 of the main text). Overall, our measurements cannot be explained by the double-well potential model with neither a narrow nor a broad defect characteristics distribution.

\begin{figure}[h]
\begin{center}
\includegraphics[width=0.31\linewidth]{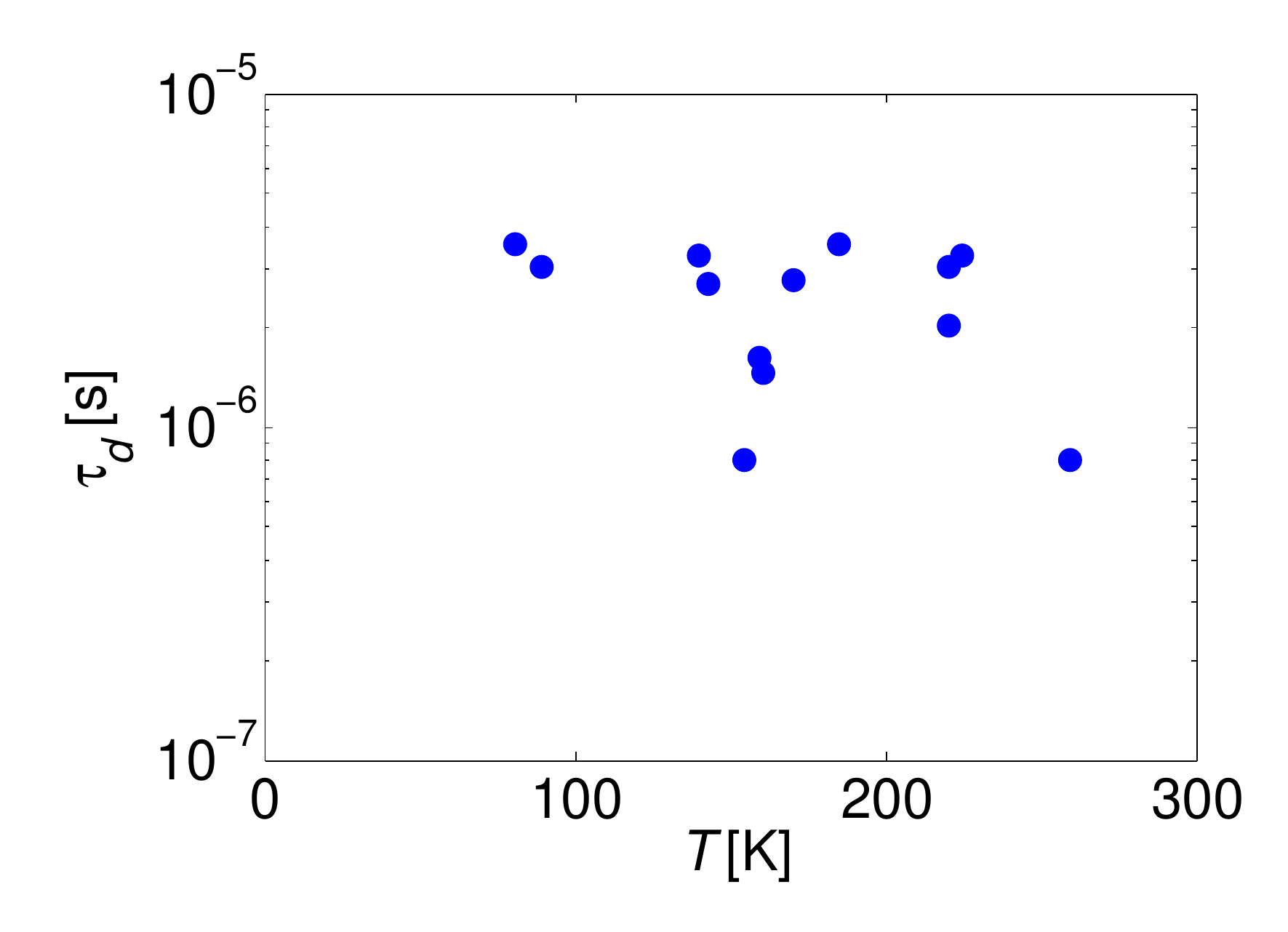}
\includegraphics[width=0.30\linewidth]{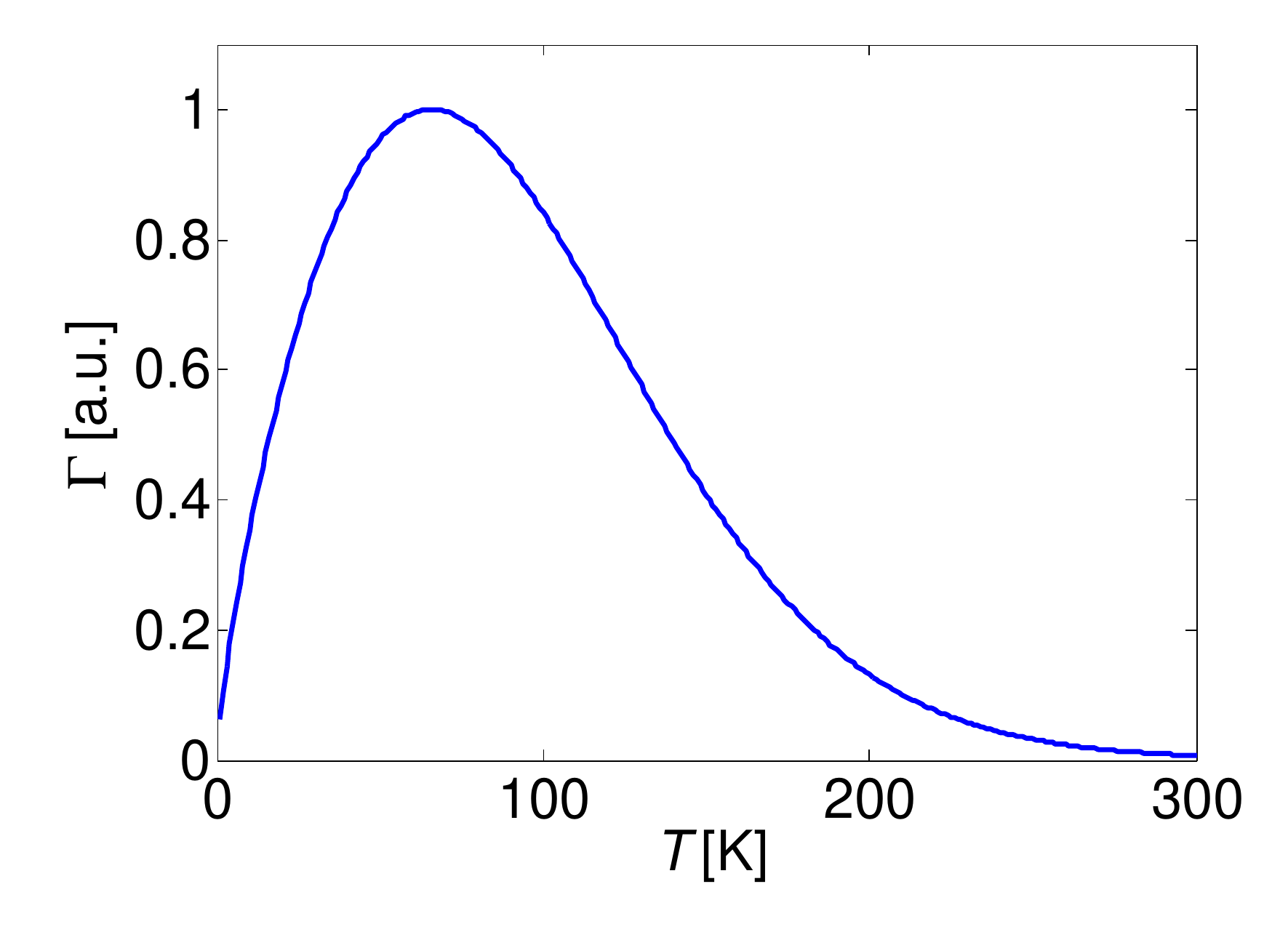}
\includegraphics[width=0.31\linewidth]{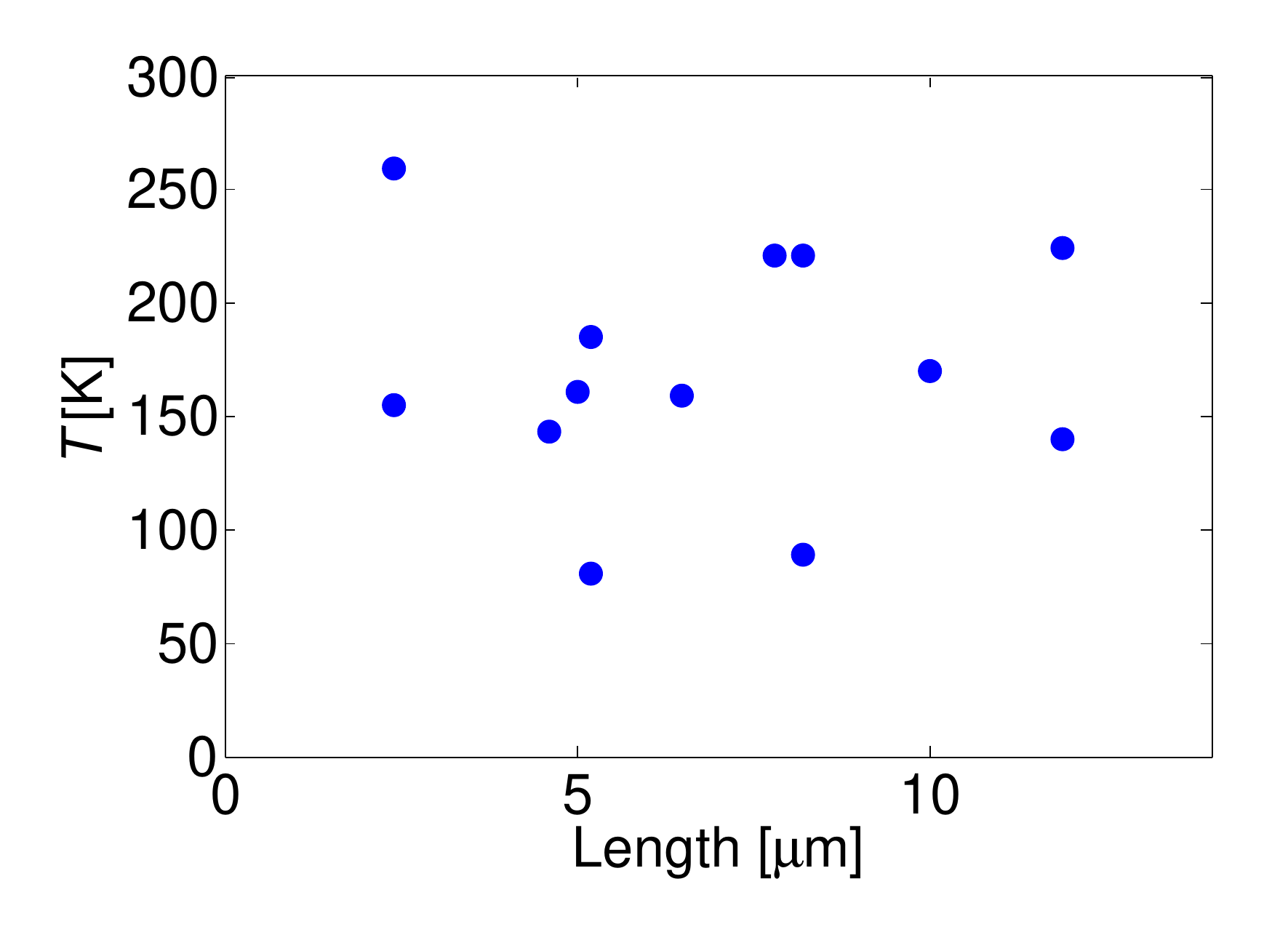}
\vspace{-.65cm}
\caption{(a) Characteristic time $\tau_d$ to overcome the barrier height as a function of temperature, obtained from all the measured devices as explained in the text. (b) Mechanical dissipation due to a distribution of defects as a function of temperature, calculated using Eq. A1 from reference \cite{SMVacher2005} with $\tau_{d0}=\SI{e-12}{\second}$, $V_0=\SI{100}{\milli\electronvolt}$, $\Delta=\SI{20}{\milli\electronvolt}$, $\omega_0/2\pi=\SI{100}{\kilo\hertz}$ and $\zeta=0.25$. (c) The temperature of the dissipation peaks as a function of the nanotube cantilever length. The data do not indicate any correlation between the dissipation peak temperature and the length. The same is observed for other geometric properties of the nanotubes, such as radius, number of walls, and radius/length ratio (not shown).}
\label{fig:TLS}
\end{center}
\end{figure}

\end{bibunit}

\end{widetext}

\end{document}